\documentclass[]{spie}

\usepackage{amsmath,amsfonts,amssymb}
\usepackage{graphicx}
\usepackage[colorlinks=true, allcolors=blue]{hyperref}

\title{Comprehensive Machine Learning Benchmarking for Fringe Projection Profilometry with Photorealistic Synthetic Data}

\author[a,*]{Anush Lakshman S}
\author[a,*]{Adam Haroon}
\author[b]{Beiwen Li}

\affil[a]{Department of Mechanical Engineering, Iowa State University, Ames, Iowa, USA}
\affil[b]{College of Engineering, University of Georgia, Athens, Georgia, USA}

\affil[*]{These authors contributed equally.}

\authorinfo{Send correspondence to Adam Haroon: E-mail: aharoon@iastate.edu}

\begin{document} 
\maketitle

\begin{abstract}

Machine learning approaches for fringe projection profilometry (FPP) are hindered by the lack of large, diverse datasets and comprehensive benchmarking protocols. This paper introduces the first open-source, photorealistic synthetic dataset for FPP, generated using NVIDIA Isaac Sim, comprising 15,600 fringe images and 300 depth reconstructions across 50 diverse objects. We apply this dataset to single-shot fringe projection profilometry, where models predict 3D depth maps directly from individual fringe images without temporal phase shifting. Through systematic ablation studies, we establish optimal learning configurations for long-range (1.5-2.1~m) depth prediction. We first compare three data normalization strategies (raw, global normalized, individual normalized), finding that individual normalization, which decouples object shape from absolute scale, achieves 9.1$\times$ better object reconstruction (16.20~mm vs 148.07~mm MAE) than raw depth. Our experiments also revealed that removing background fringe patterns catastrophically degrades performance by 2.8-7.3$\times$ across all normalizations, demonstrating that background fringes provide essential spatial phase reference rather than acting as noise. We then evaluate six loss functions (L1, MaskedL1, HybridL1, RMSE, MaskedRMSE, HybridRMSE), identifying Hybrid L1 loss ($\alpha$=0.7) as optimal, improving object error by 10\% over baseline RMSE (14.54~mm vs 16.20~mm MAE). Using the optimal configuration (individual normalization + Hybrid L1 $\alpha$=0.7), we benchmark four architectures: UNet achieves best performance (14.54~mm object MAE, 17.88~mm object RMSE), outperforming Hformer by 52\%, ResUNet by 60\%, and Pix2Pix by 91\%. Critically, Pix2Pix's counterintuitive failure reveals fundamental misalignment between adversarial training and metrological objectives. Even the best model's 14.54~mm error represents 18\% of object with an 80~mm depth range, far from FPP's sub-millimeter capabilities. The modest 1.9$\times$ performance gap between best and worst architectures confirms that the fundamental limitation is information deficit, not model design: single fringe images lack sufficient information for accurate depth recovery without explicit phase information. This work provides standardized evaluation protocols, a robust benchmark dataset, and systematic evidence that direct fringe-to-depth mapping is fundamentally limited, motivating hybrid approaches combining traditional phase-based FPP with learned refinement. Our dataset is available at \url{https://huggingface.co/datasets/aharoon/fpp-ml-bench}, and our code is available at \url{https://github.com/AnushLak/fpp-ml-bench}.

\end{abstract}


\section{Introduction and Related Work}
\label{sec:intro}

Fringe projection profilometry (FPP) has emerged as a critical non-destructive technology in robotic scanning~\cite{haroon2024autonomous}, manufacturing inspection~\cite{lakshman2024corrosion}, and 3D printing optimization~\cite{lakshman2025characterizing}, offering high-precision surface measurements with submillimeter accuracy~\cite{zhang2016high, geng2011structured}. While traditional FPP uses multi-step phase-shifting algorithms requiring sequential pattern capture, machine and deep learning (ML and DL) offers possibilities for single-shot reconstruction enabling real-time applications~\cite{zuo2022deep, zhang2010recent, zheng2020fringe, zhu2022hformer, wang2021single, ikeda2025deep, balasubramaniam2023single}.

These approaches have shown promise in phase unwrapping~\cite{wang2022deepspatialphase}, fringe denoising~\cite{yan2019fringedenoise}, and depth regression~\cite{zuo2022deep}. However, progress is hindered by three fundamental challenges. First, unlike computer vision benchmarks such as ImageNet~\cite{deng2009imagenet} or COCO~\cite{lin2014microsoft}, FPP lacks large-scale datasets with perfect ground truth and standardized evaluation protocols. Moreover, obtaining perfect 3D ground truth geometry for model training remains challenging as measurement systems introduce their own errors. To address these issues, synthetic data generation through virtual twins has proven powerful for optical metrology~\cite{nikolenko2021synthetic, de2022next} using Blender~\cite{zheng2020fringe}, Unity~\cite{ueda2021fringe}, or MATLAB~\cite{zhang2023measurement}. But these systems either require pre-calibrated physical setups or provide simplified optical models that do not reflect the complexities of real-world systems, such as over-saturation, occlusion, and subsurface scattering.

From a machine learning perspective, the field lacks systematic ablation studies to establish best design practices for single-shot FPP. Each paper employs different data representations, some normalize depth to [0,1]~\cite{feng2021calibration}, others use raw millimeter values~\cite{ikeda2025deep}, while some convert to meters~\cite{nguyen2020single}, without justifying these choices or evaluating alternatives. Similarly, loss functions vary arbitrarily across studies: L1~\cite{wang2025end}, L2~\cite{ikeda2025deep}, cross-entropy~\cite{li2025deep}, SSIM~\cite{wang2021single}, or hybrid combinations~\cite{zhu2022hformer}, with no systematic comparison of their effectiveness. This inconsistency makes it difficult to determine whether performance differences stem from architectural innovations, data preprocessing choices, or simply hyperparameter tuning. Hence, there is a need for comprehensive ablation studies to isolate the impact of each design decision. This will help the community to clearly establish whether observed errors in prediction are fundamental constraints of single-shot reconstruction or merely suboptimal configurations.

To address these issues, we build on VIRTUS-FPP~\cite{HaroonVIRTUS2025}, which introduced the first physics-based virtual FPP system with end-to-end camera-projector modeling in NVIDIA Isaac Sim, to present a systematic machine learning benchmarking framework. Our contributions include:

\begin{itemize}
    \item First open-source, photorealistic synthetic dataset for FPP, comprising 15,600 fringe images and 300 depth reconstructions across 50 diverse objects with perfect ground truth geometry at 1.5-2.1~m measurement range
    
    \item Comprehensive three-phase ablation study establishing optimal learning configurations:
    \begin{itemize}
        \item \textbf{Phase 1 - Normalization}: Individual normalization achieves 9.1$\times$ improvement over raw depth (16.20~mm vs 148.07~mm object MAE) by decoupling shape from scale
        \item \textbf{Phase 1b - Background ablation}: Counterintuitive finding that removing background fringes degrades performance 2.8-7.3$\times$, proving background provides essential spatial phase reference
        \item \textbf{Phase 2 - Loss functions}: Hybrid L1 loss ($\alpha$=0.7) achieves 10\% improvement (14.54~mm object MAE), while masked losses catastrophically fail due to scale drift
        \item \textbf{Phase 3 - Architectures}: UNet outperforms Hformer (52\%), ResUNet (60\%), and Pix2Pix (91\%) with optimal configuration
    \end{itemize}
    
    
    \item Empirical evidence that reconstruction errors (14.54-27.73~mm object MAE, representing 18-35\% of objects with an 80~mm depth range) expose fundamental limitations of direct fringe-to-depth mapping without explicit phase information: networks learn coarse shape priors rather than precise geometry. This causes a 1.9$\times$ performance gap across diverse architectures confirming information deficit, not model design, as the primary constraint
    
    \item Standardized benchmarking protocols and open-source dataset enabling systematic comparison and advancement of learning-based FPP approaches
\end{itemize}

\section{Virtual Fringe Projection Profilometry}
\label{sec:virtual}

The VIRTUS-FPP~\cite{HaroonVIRTUS2025} used for benchmarking, is built in NVIDIA Isaac Sim, integrating OptiX ray tracing for or photorealistic rendering, PhysX for physics, and Universal Scene Description (USD) for 3D composition. This section is structured as follows: Section~\ref{sec:system_config} discusses the configuration of the FPP system used for data acquisition and Section~\ref{sec:virtual_calib} elaborates the calibration process of the constructed virtual system.


\subsection{System Configuration} \label{sec:system_config}

The virtual system consists of a calibrated camera-projector pair (Table~\ref{tab:system_params}). The camera uses Isaac Sim's pinhole primitive (960$\times$960 resolution, 50 cm focal length), while the projector is modeled using a rectangular light source (0.625 m $\times$ 0.5 m, 40 nits) with texture projection. The projector is positioned 0.1 m below and 0.125 m left of the camera for optimal triangulation geometry.

\begin{table}[ht]
\caption{Virtual Camera and Projector System Parameters}
\label{tab:system_params}
\begin{center}
\begin{tabular}{|l|l|}
\hline
\rule[-1ex]{0pt}{3.5ex} \textbf{Camera Parameters} & \textbf{Value} \\
\hline
\rule[-1ex]{0pt}{3.5ex} Focal Length & 50 cm \\
\hline
\rule[-1ex]{0pt}{3.5ex} Horizontal Aperture & 20.9995 cm \\
\hline
\rule[-1ex]{0pt}{3.5ex} Vertical Aperture & 15.2908 cm \\
\hline
\rule[-1ex]{0pt}{3.5ex} Resolution & 960 $\times$ 960 pixels \\
\hline
\rule[-1ex]{0pt}{3.5ex} \textbf{Projector Parameters} & \textbf{Value} \\
\hline
\rule[-1ex]{0pt}{3.5ex} Intensity & 40 nits \\
\hline
\rule[-1ex]{0pt}{3.5ex} Height & 0.625 m \\
\hline
\rule[-1ex]{0pt}{3.5ex} Width & 0.5 m \\
\hline
\rule[-1ex]{0pt}{3.5ex} Pattern Resolution & 912 $\times$ 1140 pixels \\
\hline
\end{tabular}
\end{center}
\end{table}

VIRTUS-FPP's key innovation is projector modeling through the inverse camera model:

\begin{equation}
\begin{bmatrix} X \\ Y \\ Z \end{bmatrix} = (M_{ext})^{-1}(M_{int})^{-1}\begin{bmatrix} u \\ v \\ 1 \end{bmatrix}
\end{equation}

enabling accurate dimensional correspondence of projected fringe patterns at any distance without hardware constraints. All objects in our dataset use consistent matte material properties (roughness=0.95, specular=0.15, AO-to-diffuse=0.95) representative of typical structured light scanning~\cite{ma16155443, mi13101607}.

The rendering pipeline uses OptiX path tracing with specific configurations: disabled sampled direct lighting mode to prevent phase map artifacts, and disabled shadows for clean fringe patterns. This physics-based approach captures complex light transport including multi-bounce illumination, surface reflectivity variations, and ambient occlusion.

\subsection{Virtual Calibration} \label{sec:virtual_calib}

\begin{figure}[ht]
    \centering
    \includegraphics[width=1\textwidth]{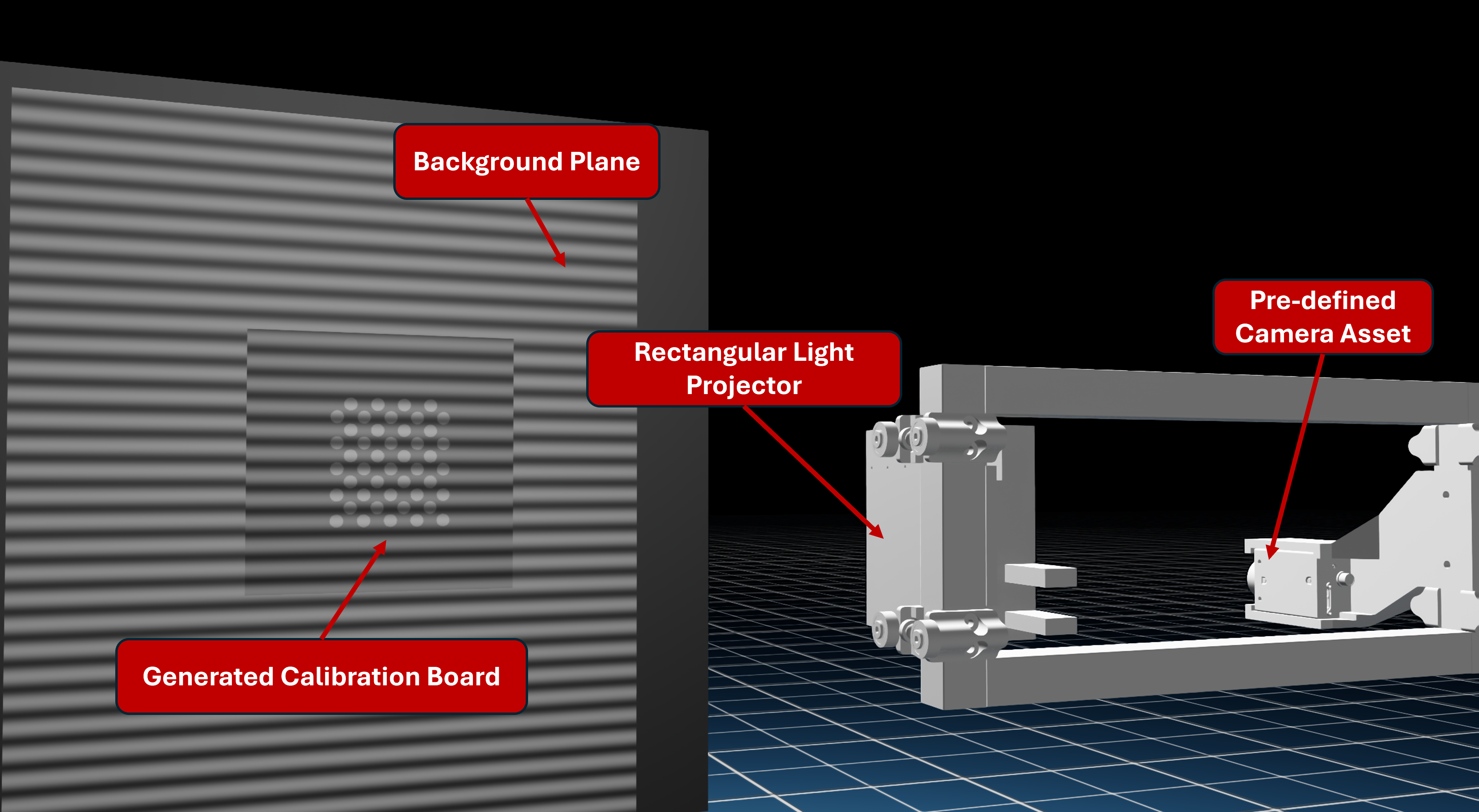}
    \caption{Virtual camera–projector calibration setup with a pinhole camera model, rectangular light-source projector, calibration board, and matte background plane.}
    \label{fig:simfpp}
\end{figure}

VIRTUS-FPP performs complete virtual calibration using procedurally generated 5$\times$9 asymmetric circular boards (10 mm diameter, 20 mm spacing). The system captures 18 calibration poses yielding 936 calibration images in ~5 minutes (10,530 images/hour). The calibrated system achieves sub-pixel accuracy (stereo reprojection error: 0.055506 pixels, projector error: 0.048609 pixels)~\cite{HaroonVIRTUS2025}. Our VIRTUS-FPP simulation setup is illustrated in Figure \ref{fig:simfpp}.

\section{Data Acquisition Methodology}
\label{sec:data}

\subsection{Dataset Composition}

We collected data for 50 USD objects from YCB datasets~\cite{calli2017yale} and NVIDIA Physical AI Warehouse~\cite{Nvidia2025PhysicalAIWarehouse} spanning cylindrical containers, rectangular boxes, complex shapes (power drills, sprayguns), and industrial components. This diversity evaluates robustness across varying surface characteristics and morphological complexity from simple geometric primitives to intricate shapes with concavities and fine-scale features. 

The system was calibrated and operated within a range of 1.5-2.1~m, with all objects positioned and scanned at distances within this range. Objects were placed on a background plane with identical matte properties to provide consistent lighting conditions and minimize unwanted reflections. For multi-view acquisition, each object was rotated about the vertical axis in 60° increments, using the rotation matrix given in Equation~\ref{eqn:rotationmatrix}:


\begin{equation}\label{eqn:rotationmatrix}
R_z(\theta_i) = \begin{bmatrix}
\cos\theta_i & -\sin\theta_i & 0 \\
\sin\theta_i & \cos\theta_i & 0 \\
0 & 0 & 1
\end{bmatrix}
\end{equation}

for $\theta_i = i \cdot 60°$ where $i = 0,1,...,5$. This yields 6 viewpoints per object with approximately 50\% overlap between adjacent views.

\subsection{Fringe Acquisition and Ground Truth Generation}
At each viewpoint, an 18-step phase-shifting sequence ($\delta_n = 2\pi n/18$, $n = 0,\ldots,17$) is captured at $960\times960$ resolution~\cite{zhang2016high}:
\begin{equation}
I_n(u,v) = I'(u,v) + I''(u,v)\cos\left[\phi(u,v) + \frac{2\pi n}{18}\right]
\end{equation}
where $(u,v)$ are pixel coordinates, $I'(u,v)$ is the background intensity, $I''(u,v)$ is the modulation amplitude, and $\phi(u,v)$ is the wrapped phase. The GPU-accelerated pipeline achieves approximately 3~fps, over twice the speed of previous approaches~\cite{zheng2020fringe}.

The captured fringe patterns are processed using the standard 18-step phase-shifting algorithm combined with Gray-code temporal unwrapping~\cite{sansoni1999three} and triangulation to generate depth maps $D(u,v)$ stored as MATLAB .mat files containing depth values in millimeters at each pixel location.

\subsection{Dataset Summary} \label{sec:dataset_summary}
The dataset comprises 15,600 raw fringe images (50 objects $\times$ 6 viewpoints $\times$ 52 fringe patterns per sequence), 300 corresponding ground truth depth maps stored as .mat files, and 50 ground truth mesh geometries. Data are partitioned with an 80/10/10 split at the object level: 240 training samples (40 objects $\times$ 6 viewpoints), 30 validation samples (5 objects $\times$ 6 viewpoints), and 30 test samples (5 objects $\times$ 6 viewpoints), ensuring evaluation on completely unseen object geometries.

For all experiments, we use the first fringe image from each 18-step sequence as the network input $I(u,v)$, ensuring identical conditions across all models. To investigate how data representation affects learning performance, we train models using three different normalization strategies for the ground truth depth maps:

\begin{enumerate}
    \item \textbf{Raw Depth (Unnormalized):} The depth values $D(u,v)$ are used directly in millimeters as stored in the .mat files, with depth values typically ranging from 0~mm (background) to 1500-2000~mm (object surfaces). This approach requires the network to learn absolute metric depth values across a large dynamic range.
    
    \item \textbf{Global Normalized Depth:} The raw depth map is normalized by a global constant, converting millimeters to meters~\cite{nguyen2020single}:
    \begin{equation}
        D_{\text{global}}(u,v) = \frac{D(u,v)}{1000}
    \end{equation}
    This strategy reduces the numerical range to approximately 0-2~meters, which may improve numerical stability during training while maintaining a consistent scale across all objects.
    
    \item \textbf{Individual Normalized Depth:} Each depth map is normalized independently to the range $[0,1]$ based on its object-specific depth range:
    \begin{equation}
        D_{\text{norm}}(u,v) = \begin{cases}
            \frac{D(u,v) - D_{\min}}{D_{\max} - D_{\min}} & \text{if } D(u,v) > 0 \\
            0 & \text{if } D(u,v) = 0
        \end{cases}
    \end{equation}
    where $D_{\min} = \min_{u,v}\{D(u,v) \mid D(u,v) > 0\}$ and $D_{\max} = \max_{u,v}\{D(u,v)\}$ are computed per object. The normalization parameters $(D_{\min}, D_{\max})$ are stored separately for each training sample to enable metric reconstruction during evaluation. This approach normalizes the learning problem to a consistent $[0,1]$ range across all training samples, effectively decoupling the learning of object \emph{shape} from absolute \emph{scale}.
\end{enumerate}

\begin{figure}[ht]
    \centering
    \includegraphics[width=0.32\linewidth]{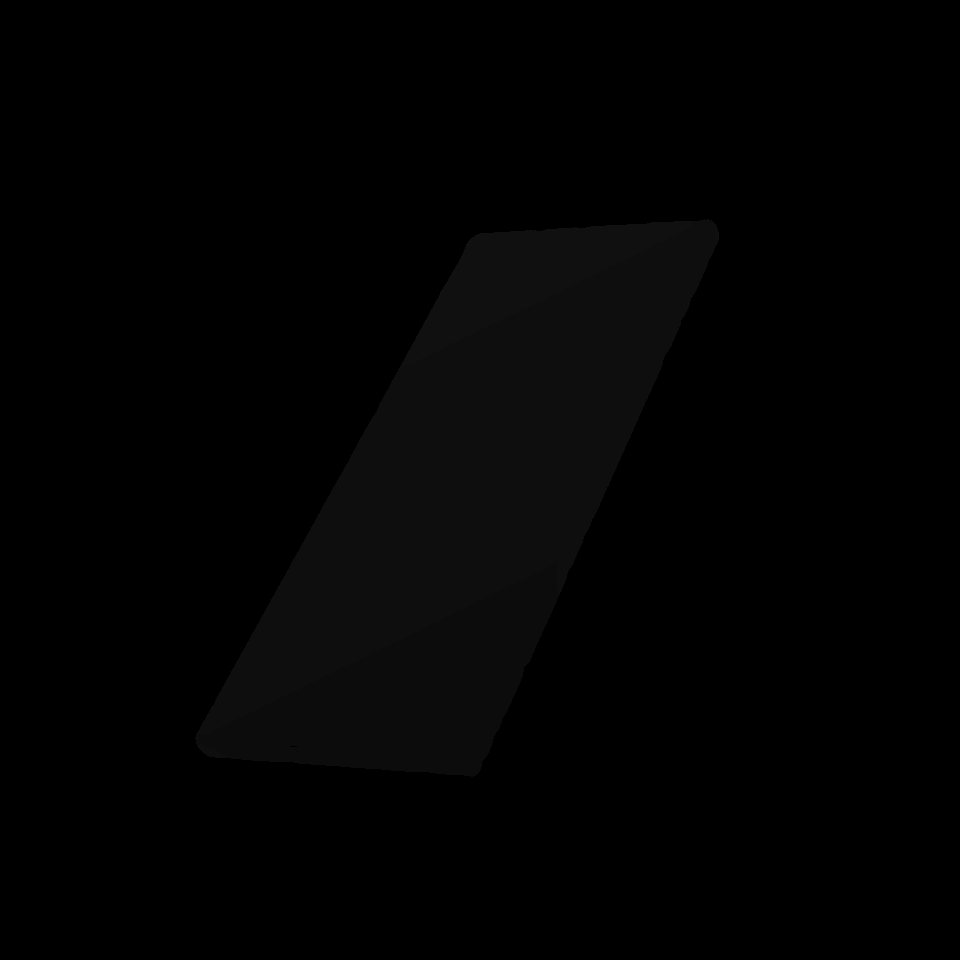}
    \includegraphics[width=0.32\linewidth]{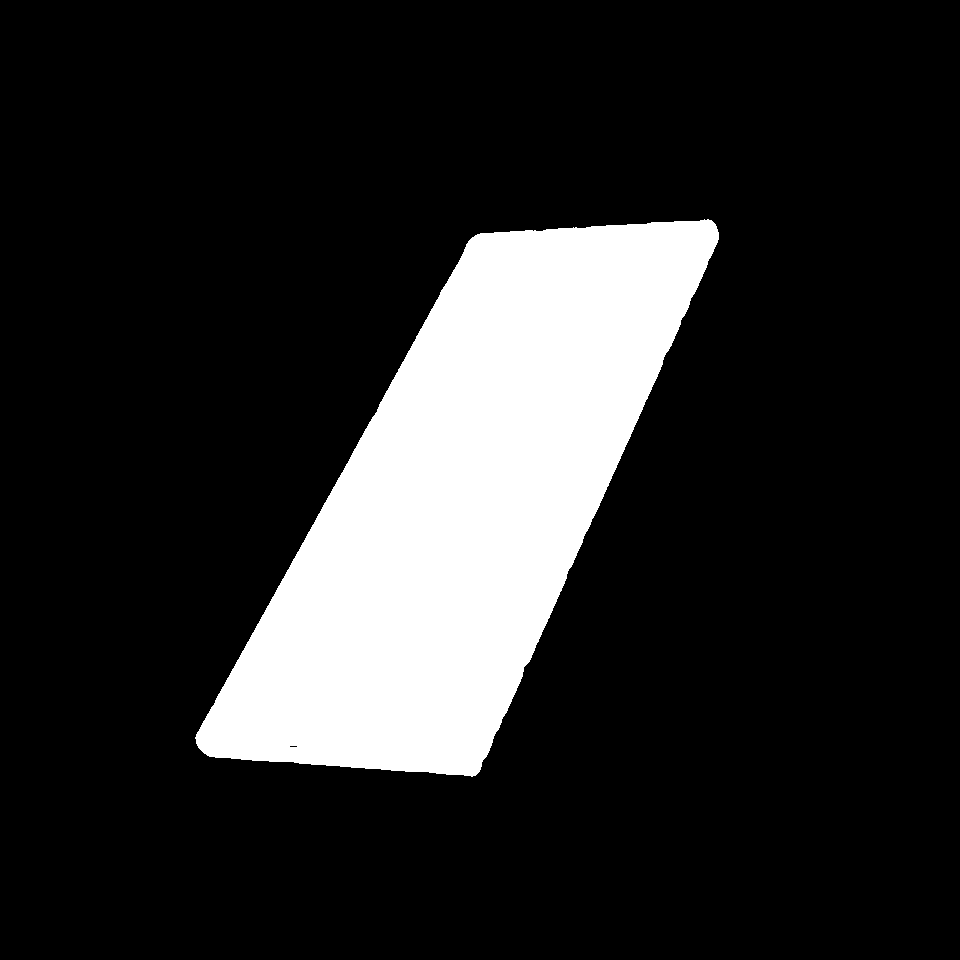}
    \includegraphics[width=0.32\linewidth]{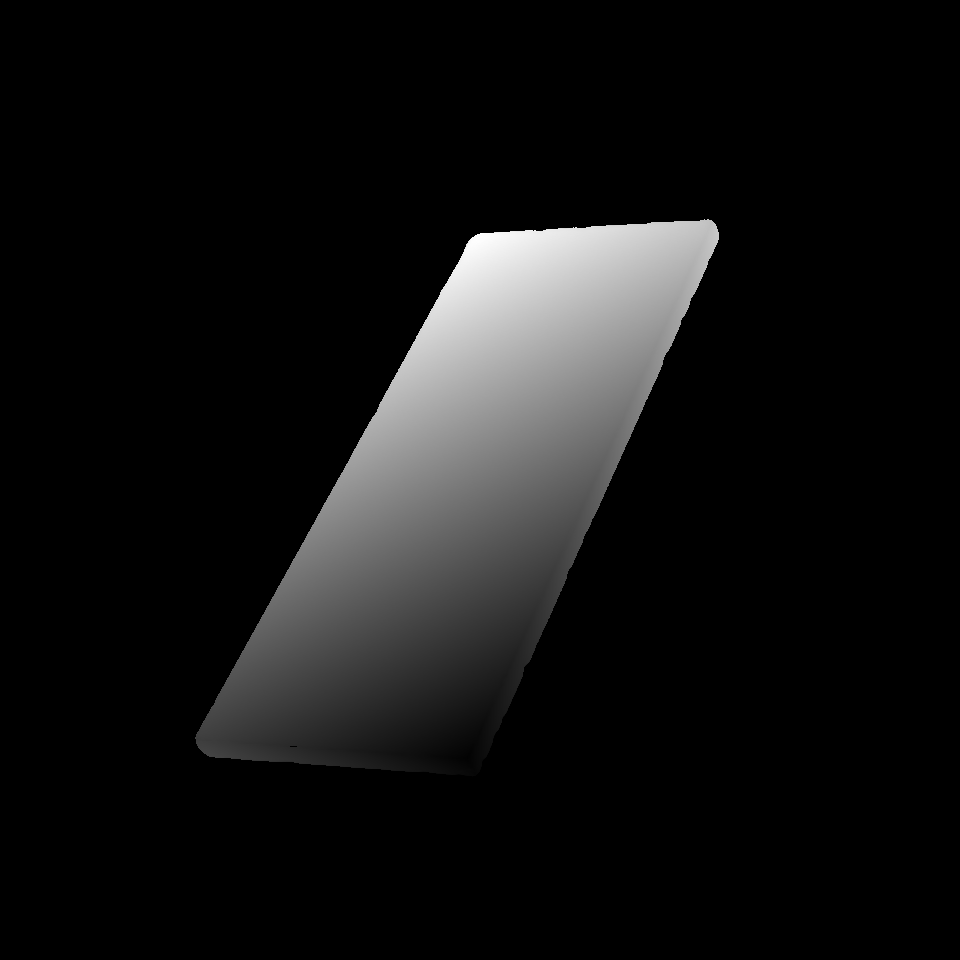}
    \caption{Depth map visualizations for three normalization strategies on the same object (wooden boards). From left to right: raw depth (0-2026~mm), global normalized depth (0-2.026~m), and individual normalized depth ($[0,1]$ range mapped to 1561-2026~mm). All use the same underlying depth data, differing only in normalization.}
    \label{fig:depth_normalizations}
\end{figure}

Figure~\ref{fig:depth_normalizations} shows example depth visualizations for each normalization strategy. While all three normalizations use the same underlying .mat depth files for training, we visualize these normalizations as depth images by converting them to uint16 format. This conversion is performed solely for visualization purposes: given a depth map in any normalization, we apply min-max scaling to directly map the depth values to the uint16 range $[0, 65535]$. The actual training data remain as floating-point depth values in .mat format.

\section{Single-Shot 3D Reconstruction Benchmarking}

\subsection{Problem Formulation}

Single-shot reconstruction predicts depth $\hat{D}_{\text{norm}} = f_\theta(I)$ from a single fringe image $I$, where $f_\theta$ is a neural network. We use the first fringe from each 18-step sequence as input, ensuring identical conditions across models. This task is inherently challenging, as single fringe images contain ambiguity in depth estimation due to the periodic nature of sinusoidal patterns. In the absence of temporal dependency or spatial unwrapping, each fringe cycle spans a $2\pi$ phase range, making it difficult to uniquely associate a specific cycle with a surface point. As a result, learning-based approaches rely on inferring depth from learned shape priors and statistical regularities, rather than from fully explicit geometric cues alone.

To systematically evaluate this challenging problem, we structure our investigation around three key research questions: (1) How does depth data normalization affect model performance? (2) Which loss functions are most effective for this task? (3) Which network architectures achieve the best reconstruction accuracy? This methodology allows us to isolate the impact of each design choice and establish best practices for learning-based single-shot FPP.

\subsection{Normalization Strategy Comparison}\label{sec:normalization_comparison}

\textbf{Architecture and Training Configuration:} We use UNet~\cite{ronneberger2015} as our baseline architecture due to its widespread adoption in FPP literature~\cite{feng2021calibration,zheng2020fringe,ikeda2025deep} and strong performance on dense prediction tasks. The network consists of four encoder-decoder stages, progressively downsampling from $960\times960$ to $60\times60$ resolution, with channel depths increasing from 64 to 1024 at the bottleneck. Each encoder and decoder stage uses two $3\times3$ convolutions followed by instance normalization and ReLU activation. Skip connections preserve spatial information across corresponding encoder-decoder levels.

We train with RMSE loss using the RMSprop optimizer ($\alpha=0.99$, initial learning rate $10^{-4}$, weight decay $10^{-5}$), batch size 4, and ReduceLROnPlateau scheduler (factor=0.5, patience=10 epochs, minimum learning rate $10^{-6}$). This combination has demonstrated effectiveness for similar dense regression tasks~\cite{ikeda2025deep} and provides a stable baseline for comparison. Training continues until the learning rate reaches its minimum threshold, indicating convergence.

\textbf{Evaluation Metrics:} A critical challenge in evaluating FPP reconstruction is that background pixels (which contain no valid depth information and should ideally predict 0~mm or a background-specific value) can numerically dominate error statistics. Since background regions typically occupy 60-90\% of the image, they can significantly dilute object prediction errors, leading to misleadingly low overall error metrics. To address this, we report three complementary metrics for each normalization strategy:

\begin{itemize}
    \item \textbf{Overall Error:} Computed across all pixels, providing a holistic system-level metric:
    \begin{equation}
        \text{MAE}_{\text{overall}} = \frac{1}{HW}\sum_{u,v} |\hat{D}(u,v) - D(u,v)|
    \end{equation}
    \begin{equation}
        \text{RMSE}_{\text{overall}} = \sqrt{\frac{1}{HW}\sum_{u,v} (\hat{D}(u,v) - D(u,v))^2}
    \end{equation}
    
    \item \textbf{Object Error:} Computed only on pixels where ground truth depth is non-zero ($D(u,v) > 0$ for raw/global normalizations, or $D_{\text{norm}}(u,v) > 0$ for individual normalization), isolating prediction accuracy on the actual object geometry:
    \begin{equation}
        \text{MAE}_{\text{object}} = \frac{\sum_{u,v} [D(u,v) > 0] \cdot |\hat{D}(u,v) - D(u,v)|}{\sum_{u,v} [D(u,v) > 0]}
    \end{equation}
    \begin{equation}
        \text{RMSE}_{\text{object}} = \sqrt{\frac{\sum_{u,v} [D(u,v) > 0] \cdot (\hat{D}(u,v) - D(u,v))^2}{\sum_{u,v}  [D(u,v) > 0]}}
    \end{equation}
    
    \item \textbf{Background Error:} Computed only on pixels where ground truth depth is zero, measuring background suppression performance:
    \begin{equation}
        \text{MAE}_{\text{bg}} = \frac{\sum_{u,v} [D(u,v) = 0] \cdot |\hat{D}(u,v) - D(u,v)|}{\sum_{u,v} [D(u,v) = 0]}
    \end{equation}
    \begin{equation}
        \text{RMSE}_{\text{bg}} = \sqrt{\frac{\sum_{u,v}  [D(u,v) = 0] \cdot (\hat{D}(u,v) - D(u,v))^2}{\sum_{u,v}  [D(u,v) = 0]}}
    \end{equation}
\end{itemize}

All errors are reported in millimeters after denormalization to enable direct comparison across normalization strategies. For individual normalized predictions, we apply the inverse transformation $\hat{D}(u,v) = \hat{D}_{\text{norm}}(u,v) \cdot (D_{\max} - D_{\min}) + D_{\min}$ using the stored normalization parameters.

\begin{figure}[ht]
    \centering
    \includegraphics[width=0.9\linewidth]{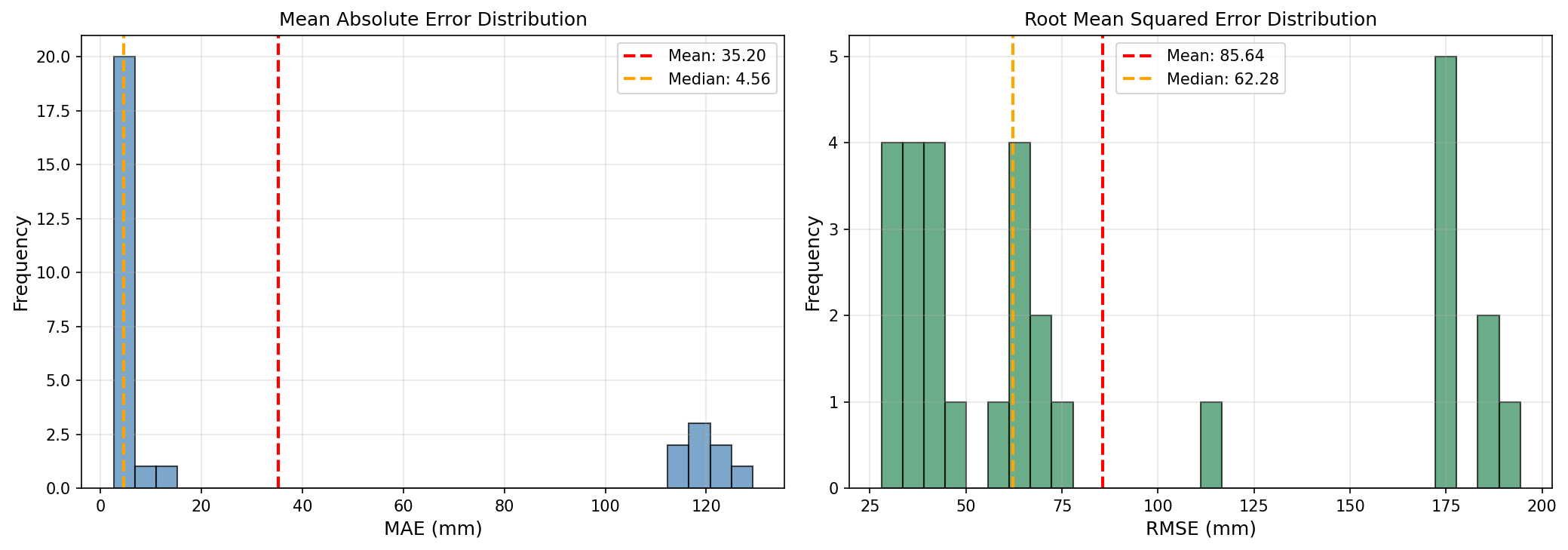}\\
    \vspace{0.3cm}
    \includegraphics[width=0.9\linewidth]{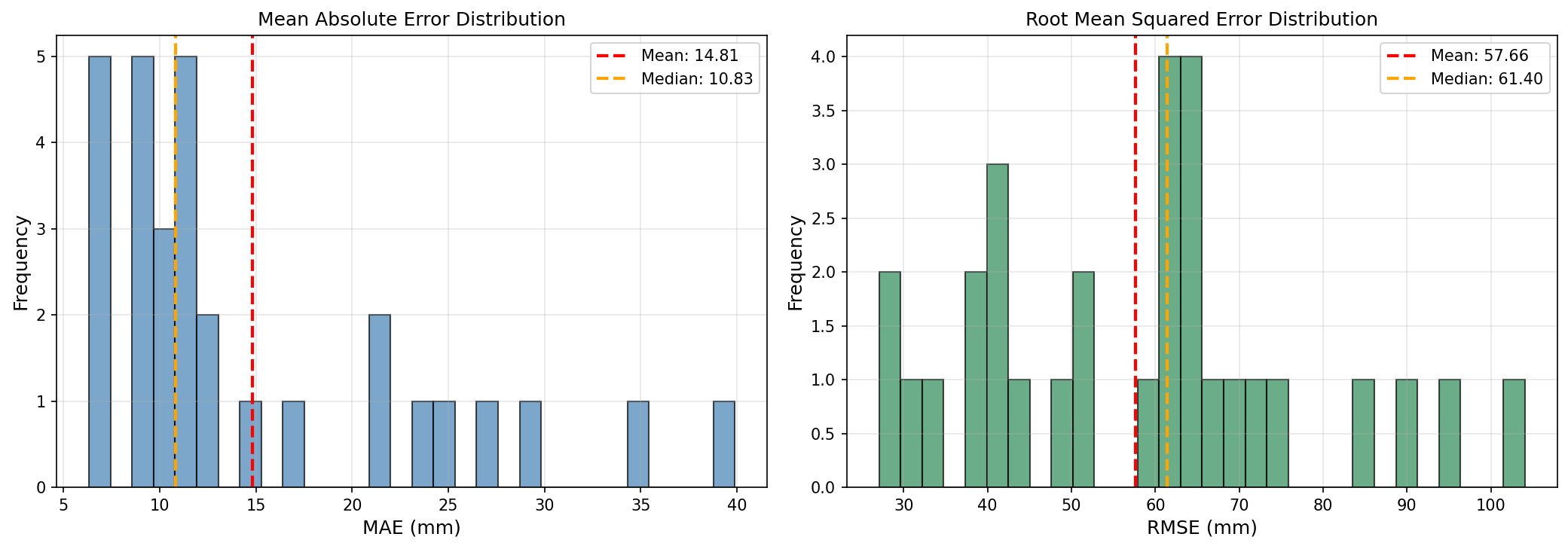}\\
    \vspace{0.3cm}
    \includegraphics[width=0.9\linewidth]{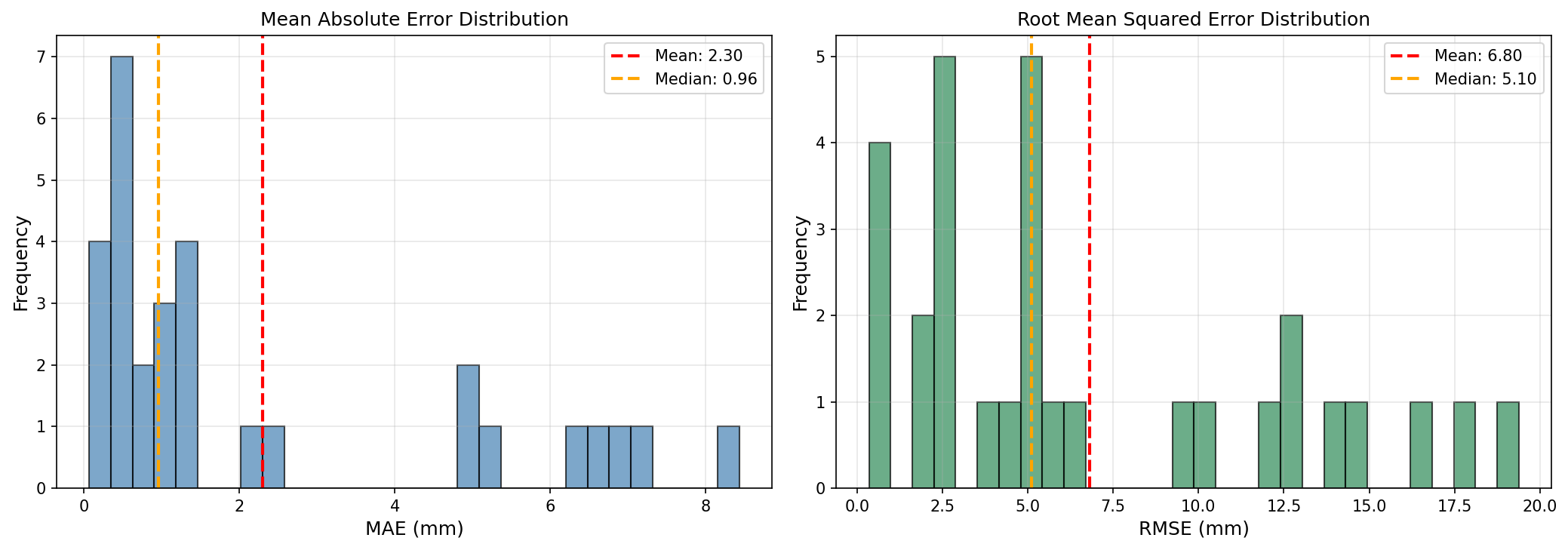}
    \caption{Overall error distributions (MAE and RMSE) across 30 test samples for three normalization strategies. Top: Raw depth shows high variance with mean MAE 35.20~mm. Middle: Global normalized depth reduces error to 14.81~mm mean MAE. Bottom: Individual normalized depth achieves 2.30~mm mean MAE with tight distribution, demonstrating superior and consistent performance.}
    \label{fig:normalization_results}
\end{figure}

\begin{table}[ht]
\caption{Reconstruction error for three normalization strategies on 30 test samples. All errors reported in millimeters.}
\label{tab:normalization_results}
\begin{center}
\begin{tabular}{|l|c|c|c|c|c|c|}
\hline
\rule[-1ex]{0pt}{3.5ex} \textbf{Normalization} & \textbf{Overall} & \textbf{Overall} & \textbf{Object} & \textbf{Object} & \textbf{Background} & \textbf{Background} \\
\rule[-1ex]{0pt}{3.5ex} & \textbf{MAE} & \textbf{RMSE} & \textbf{MAE} & \textbf{RMSE} & \textbf{MAE} & \textbf{RMSE} \\
\hline
\rule[-1ex]{0pt}{3.5ex} Raw & 35.20 & 85.64 & 148.07 & 214.37 & 19.41 & 52.38 \\
\hline
\rule[-1ex]{0pt}{3.5ex} Global & 14.81 & 57.66 & 82.49 & 144.23 & 9.90 & 45.69 \\
\hline
\rule[-1ex]{0pt}{3.5ex} Individual & \textbf{2.30} & \textbf{6.80} & \textbf{16.20} & \textbf{21.19} & \textbf{0.92} & \textbf{3.00} \\
\hline
\end{tabular}
\end{center}
\end{table}

\begin{figure}[ht]
    \centering
    \includegraphics[width=0.95\linewidth]{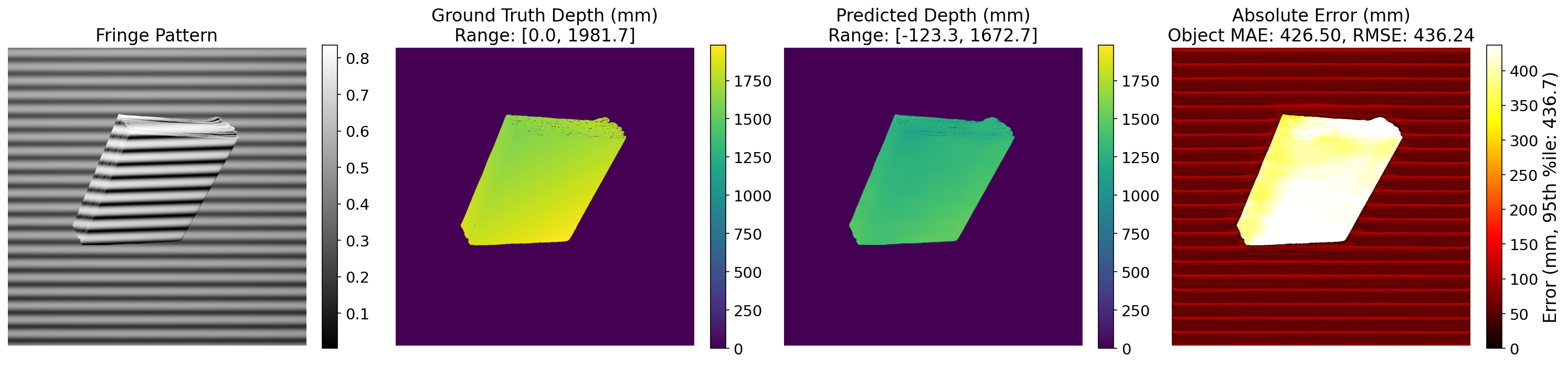}
    \vspace{0.2cm}
    \includegraphics[width=0.95\linewidth]{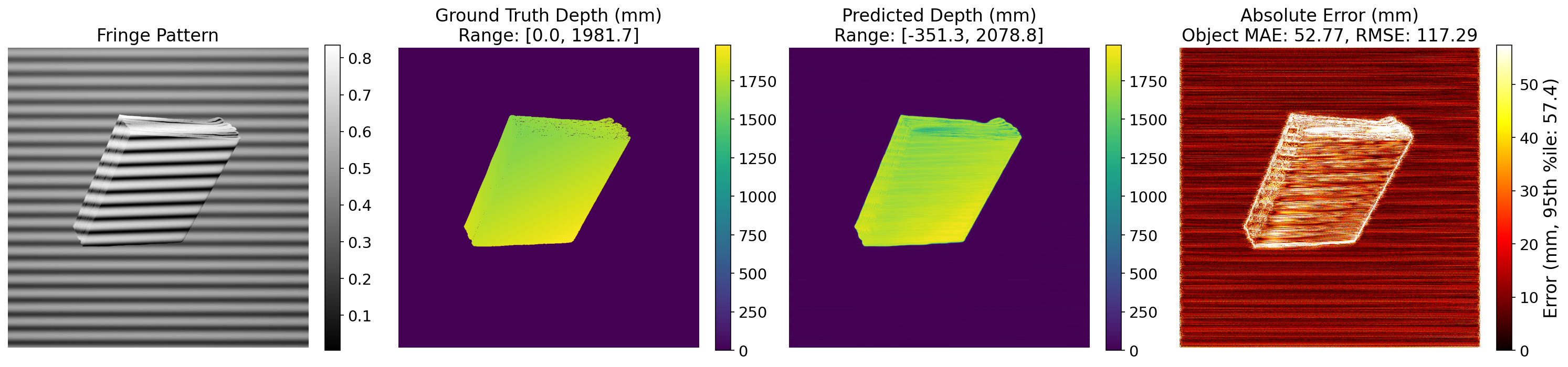}\\
    \vspace{0.2cm}
    \includegraphics[width=0.95\linewidth]{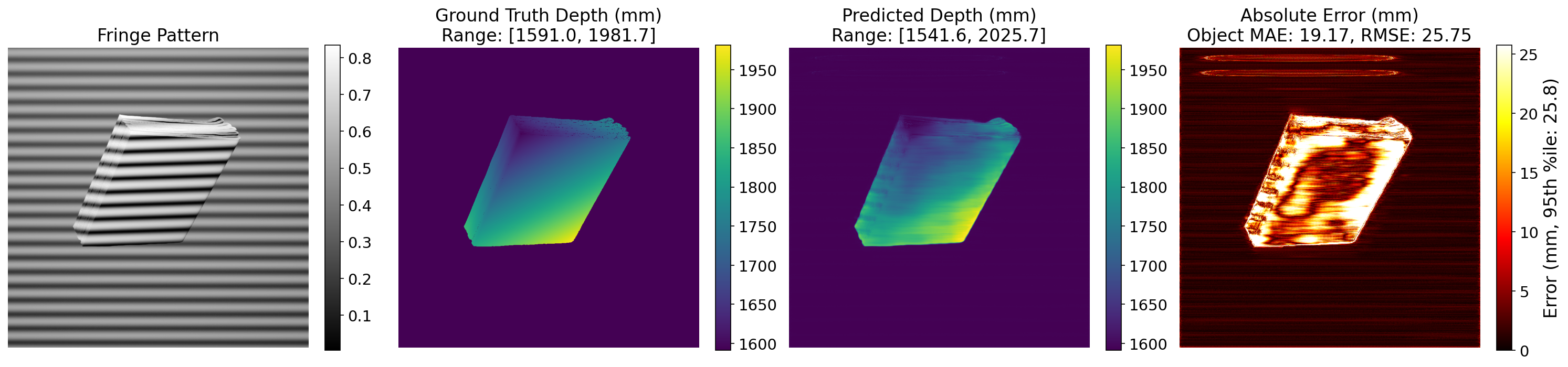}
    \caption{Single-shot depth reconstruction for magazine stack object comparing three normalization strategies. From top to bottom: raw depth (426.50~mm object MAE), global normalized depth (52.77~mm object MAE), individual normalized depth (19.17~mm object MAE). Each row from left to right shows: input fringe pattern, ground truth depth, predicted depth, absolute error map (clipped at 95th percentile for visibility). Individual normalization achieves 22× improvement over raw and 2.7× improvement over global.}
    \label{fig:normalization_predictions}
\end{figure}

\textbf{Results:} Figure~\ref{fig:normalization_results} shows overall error distributions across the 30 test samples for each normalization strategy. Figure~\ref{fig:normalization_predictions} visualizes detailed per-sample prediction for a representative object (magazine stack) under each normalization strategy. Table~\ref{tab:normalization_results} summarizes the quantitative results.

Individual normalization substantially outperforms both alternatives across all metrics. Most notably, the object-only MAE (which directly measures geometric reconstruction accuracy) improves from 148.07~mm (raw) to 82.49~mm (global) to just 16.20~mm (individual), representing a 9.1$\times$ improvement over raw depth and a 5.1$\times$ improvement over global normalization. Background suppression also improves dramatically, with individual normalization achieving near-zero background error (0.92~mm MAE, 3.00~mm RMSE) compared to 19.41~mm MAE for raw depth.

The overall error statistics can be misleading without this decomposition: raw depth shows 35.20~mm overall MAE despite catastrophic 148.07~mm object error because the numerically larger number of background pixels (with lower 19.41~mm error) dilute the overall statistic. This highlights the importance of reporting object-specific metrics for FPP evaluation.

\begin{figure}[ht]
    \centering
    \includegraphics[width=0.9\linewidth]{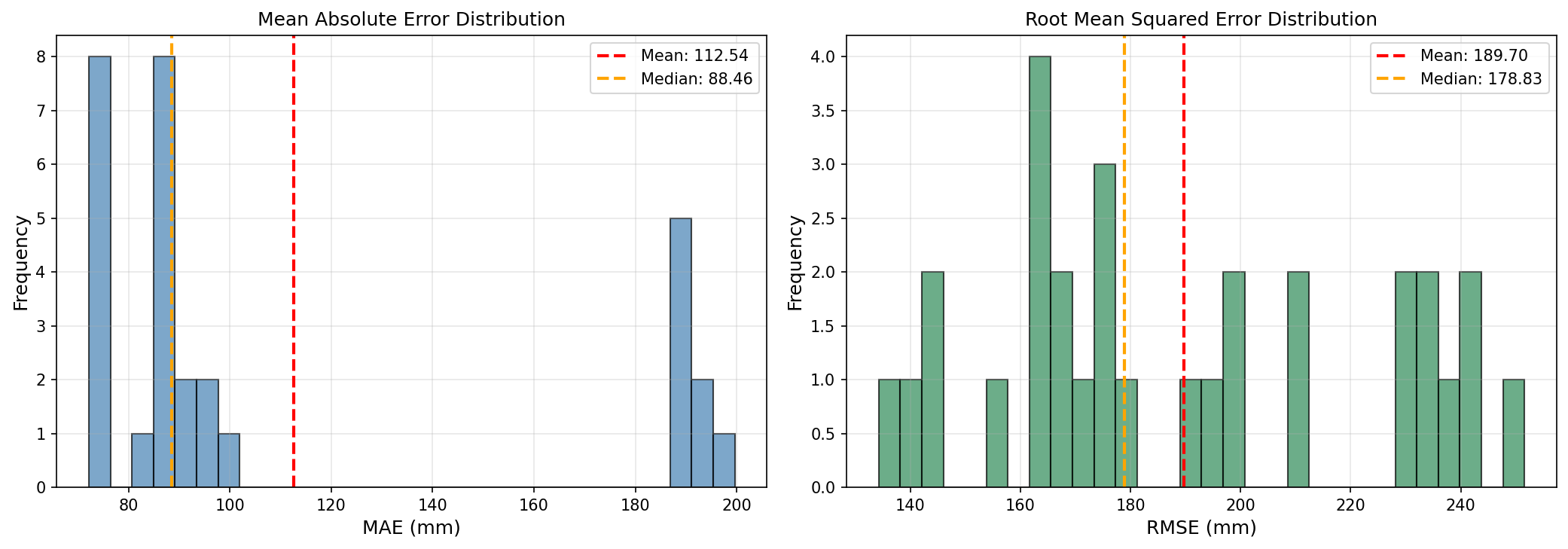}\\
    \vspace{0.3cm}
    \includegraphics[width=0.9\linewidth]{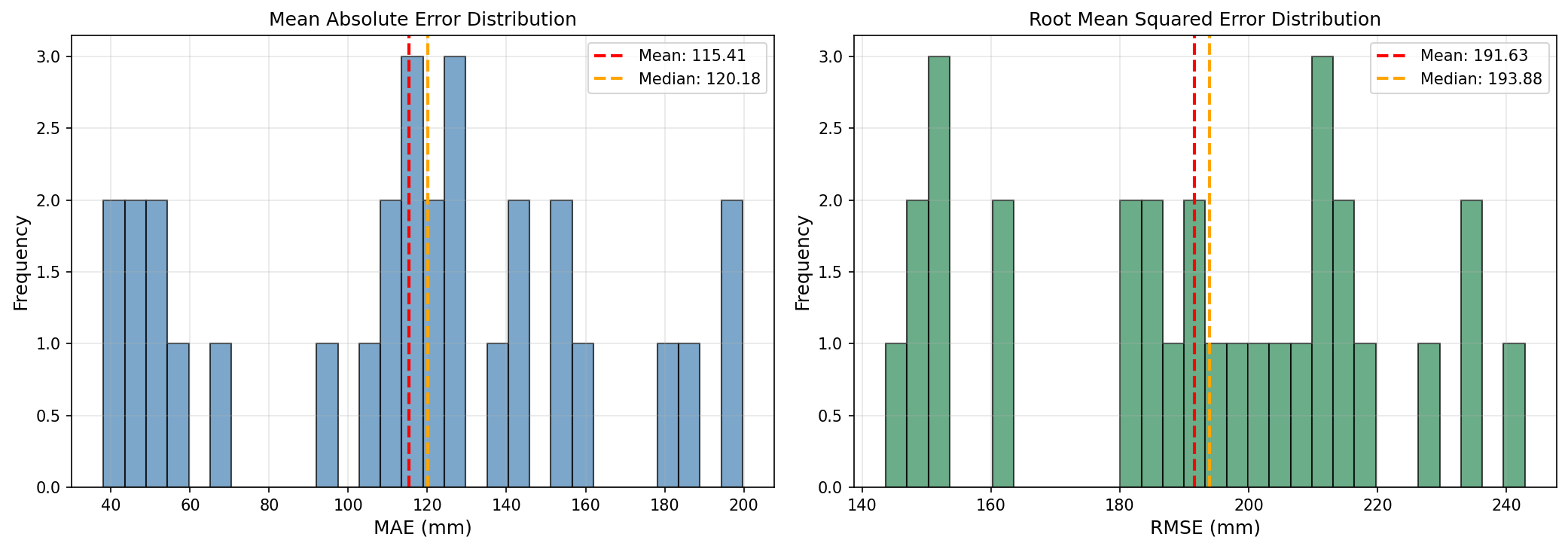}\\
    \vspace{0.3cm}
    \includegraphics[width=0.9\linewidth]{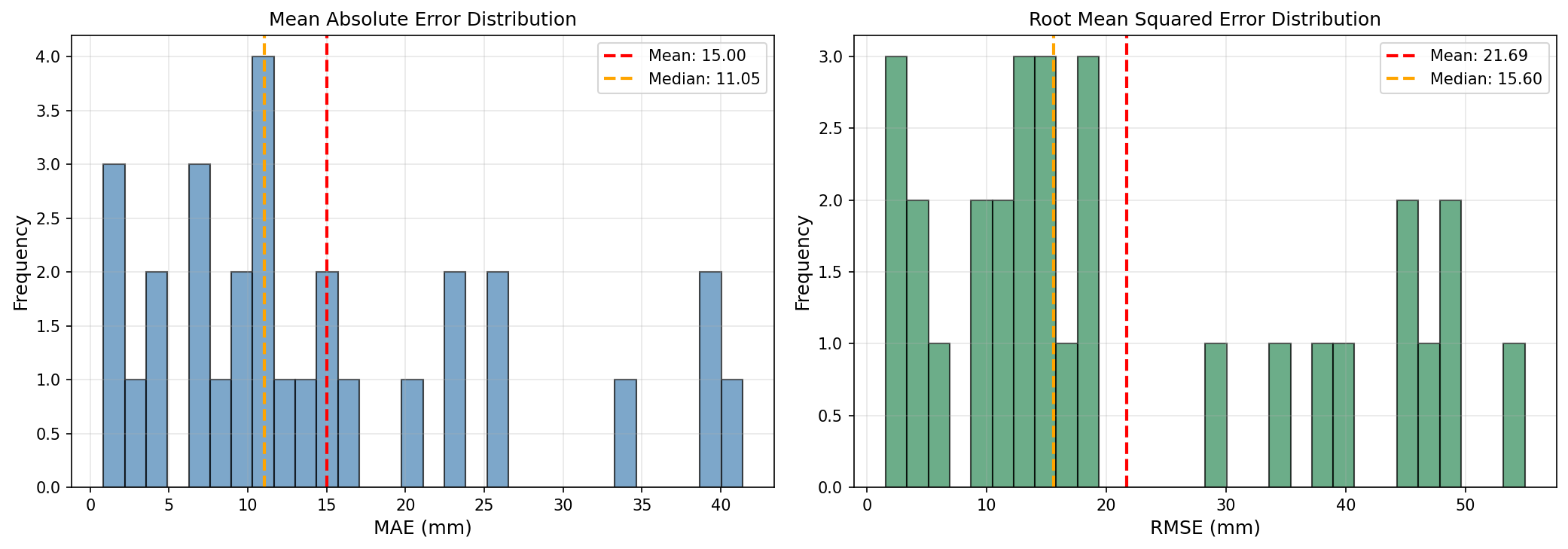}
    \caption{Overall error distributions for background-removed fringe inputs. Compare with Figure~\ref{fig:normalization_results}: all normalization strategies show dramatically increased error when background fringes are removed. Individual normalized increases from 2.30~mm to 15.00~mm mean MAE (6.5$\times$ worse).}
    \label{fig:background_ablation_results}
\end{figure}

The performance ranking directly correlates with optimization difficulty. Raw depth requires the network to learn absolute metric values across a 0-2000~mm range while simultaneously learning object geometry, creating a challenging multi-scale optimization problem. Global normalization improves this by reducing the range to 0-2~meters, but different objects still have different depth ranges in this normalized space, requiring the network to handle scale variation. Individual normalization fundamentally simplifies the learning problem by decoupling scale from shape. In the $[0,1]$ normalized space, the network need only learn the relative depth structure (the \emph{shape} of each object) with the absolute scale provided by the stored normalization parameters. This consistent $[0,1]$ range across all training samples creates a more uniform optimization landscape. Additionally, individual normalization provides an implicit regularization effect: background pixels (which map to 0.0 in normalized space for all objects) become trivial to learn, resulting in near-perfect background suppression that persists after denormalization.

\subsubsection{Background Fringe Ablation Study}\label{sec:background_ablation}

Visual inspection of prediction error maps reveals horizontal fringe artifacts in background regions of predicted depth, suggesting the network attempts to predict depth from background fringe patterns. To investigate whether these background fringes help or hinder learning, we conduct an ablation study by removing them from the input.

\textbf{Experimental Setup:} We mask the input fringe images by setting all background pixels (where $D(u,v) = 0$) to zero intensity, effectively removing the background fringe pattern. We retrain the same UNet architecture with identical hyperparameters on all three normalization strategies using these background-removed inputs, then evaluate on test data with similarly masked inputs.

\begin{table}[ht]
\caption{Impact of background fringe removal on reconstruction error (millimeters). All three normalization strategies show severe performance degradation when background fringes are removed from input images.}
\label{tab:background_ablation}
\begin{center}
\begin{tabular}{|l|c|c|c|c|c|c|}
\hline
\rule[-1ex]{0pt}{3.5ex} \textbf{Normalization} & \textbf{Overall} & \textbf{Overall} & \textbf{Object} & \textbf{Object} & \textbf{Background} & \textbf{Background} \\
\rule[-1ex]{0pt}{3.5ex} & \textbf{MAE} & \textbf{RMSE} & \textbf{MAE} & \textbf{RMSE} & \textbf{MAE} & \textbf{RMSE} \\
\hline
\rule[-1ex]{0pt}{3.5ex} Raw & 112.54 & 189.70 & 437.40 & 511.08 & 91.99 & 178.83 \\
\hline
\rule[-1ex]{0pt}{3.5ex} Global & 115.41 & 191.63 & 598.40 & 723.93 & 92.26 & 151.88 \\
\hline
\rule[-1ex]{0pt}{3.5ex} Individual & 15.00 & 21.69 & 45.01 & 57.07 & 12.16 & 15.60 \\
\hline
\end{tabular}
\end{center}
\end{table}

\begin{figure}[ht]
    \centering
    \includegraphics[width=0.95\linewidth]{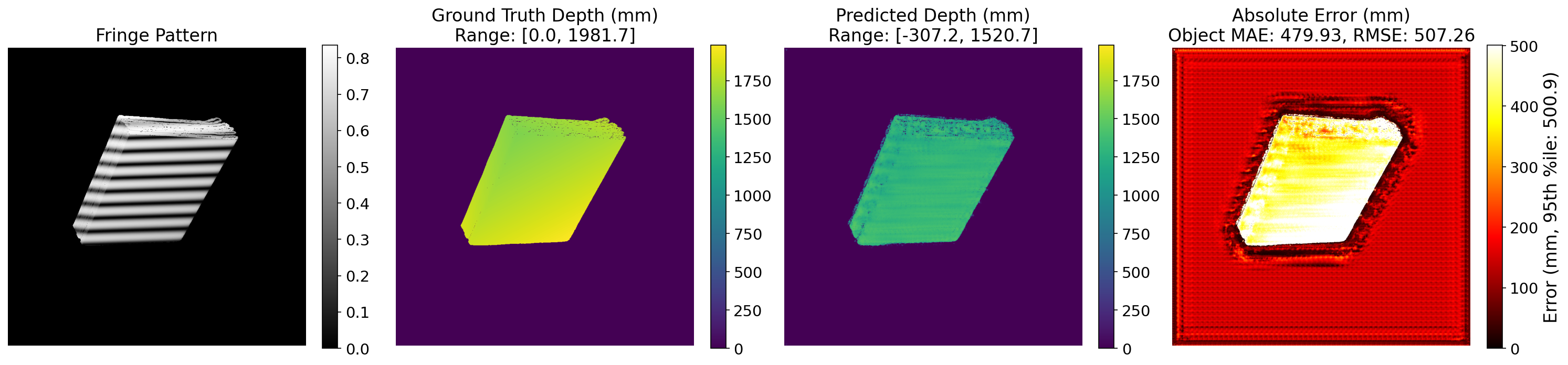}
    \vspace{0.2cm}
    \includegraphics[width=0.95\linewidth]{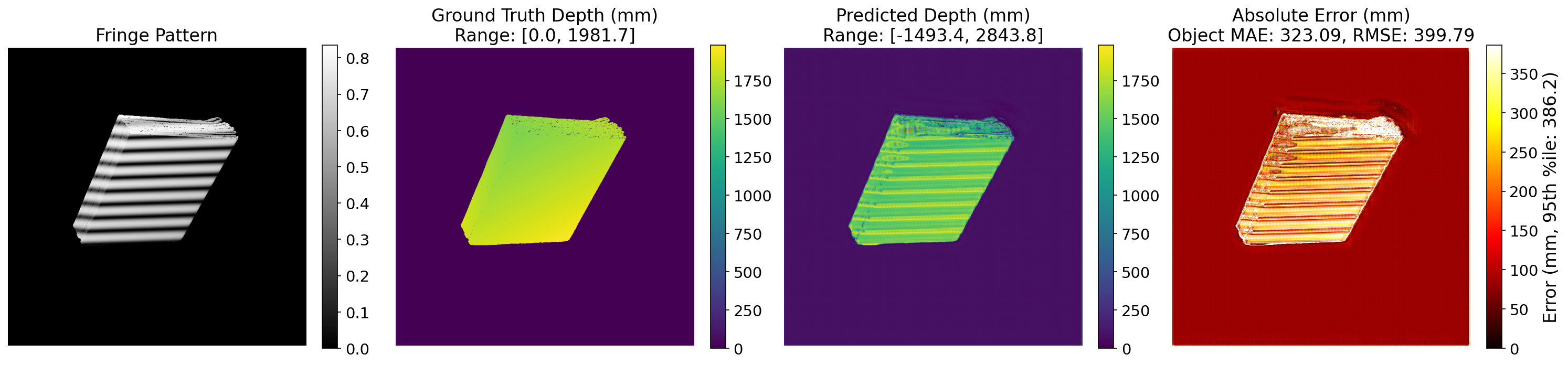}\\
    \vspace{0.2cm}
    \includegraphics[width=0.95\linewidth]{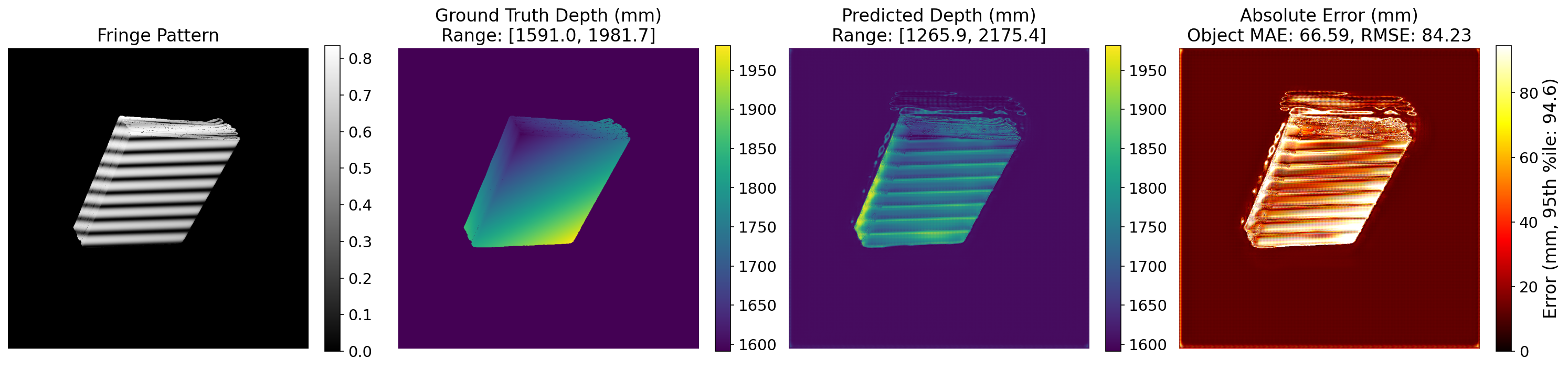}
    \caption{Impact of background fringe removal across three normalization strategies for the same magazine stack object. From top to bottom: raw depth (479.93~mm object MAE), global normalized depth (323.03~mm object MAE), individual normalized depth (66.59~mm object MAE). Background removal degrades all strategies: raw 1.1×, global 6.1×, individual 3.5× worse. Note severe boundary artifacts and inconsistent depth prediction compared to Figure~\ref{fig:normalization_predictions}.}
    \label{fig:background_ablation_predictions}
\end{figure}

\textbf{Results:} Table~\ref{tab:background_ablation} and Figure~\ref{fig:background_ablation_results} show that, surprisingly, background removal \emph{catastrophically degrades} performance across all normalization strategies. For individual normalization, object MAE increases from 16.20~mm to 45.01~mm (2.8$\times$ worse), while background error increases from 0.92~mm to 12.16~mm (13.2$\times$ worse). Raw and global normalizations suffer even more severe degradation, with object errors increasing by 3.0$\times$ and 7.3$\times$ respectively. Figure~\ref{fig:background_ablation_predictions} shows that predictions with masked inputs exhibit severe boundary artifacts and depth discontinuities not present with full fringe patterns.

This counterintuitive result reveals that background fringes provide essential spatial context rather than acting as distractors. The continuous fringe pattern across the entire image enables the network to establish a spatial phase reference that helps resolve the inherent $2\pi$ ambiguity of single-shot reconstruction. Natural phase discontinuities at object boundaries provide clearer edge information than the artificial black-to-object transitions created by masking, allowing the network to learn where fringe patterns abruptly change phase as indicators of depth transitions. Additionally, background pixels provide training signal and implicit regularization, helping the network understand the overall fringe-to-depth relationship and preventing overfitting to object-only patterns. This finding has important implications for FPP learning: background fringes are \emph{signal}, not noise. All subsequent experiments therefore use full fringe patterns with background intact.

\subsection{Loss Function Comparison}\label{sec:loss_comparison}

Having established individual normalization as the optimal data representation and confirmed that background fringes should be retained, we now investigate whether alternative loss functions can further improve reconstruction accuracy. We evaluate six L1/L2-based loss functions by training the UNet architecture with individual normalized depth and full fringe inputs:

\begin{enumerate}
    \item \textbf{RMSE Loss (Baseline):} Standard root mean squared error computed over all pixels:
    \begin{equation}
        \mathcal{L}_{\text{RMSE}} = \sqrt{\frac{1}{HW}\sum_{u=1}^{W}\sum_{v=1}^{H} (\hat{D}(u,v) - D(u,v))^2 + \epsilon}
    \end{equation}
    where $\epsilon = 10^{-8}$ ensures numerical stability.
    
    \item \textbf{L1 Loss:} Mean absolute error computed over all pixels, which is less sensitive to outliers than RMSE:
    \begin{equation}
        \mathcal{L}_{\text{L1}} = \frac{1}{HW}\sum_{u=1}^{W}\sum_{v=1}^{H} |\hat{D}(u,v) - D(u,v)|
    \end{equation}
    
    \item \textbf{Masked RMSE Loss:} RMSE computed only on valid object pixels where $D(u,v) > 0$:
    \begin{equation}
        \mathcal{L}_{\text{MaskedRMSE}} = \sqrt{\frac{\sum_{u,v} \mathcal{M}(u,v) \cdot (\hat{D}(u,v) - D(u,v))^2}{\sum_{u,v} \mathcal{M}(u,v)} + \epsilon}
    \end{equation}
    where $\mathcal{M}(u,v) =  [D(u,v) > 0]$ is a binary mask. This forces the network to focus exclusively on object geometry.
    
    \item \textbf{Masked L1 Loss:} L1 loss computed only on valid pixels:
    \begin{equation}
        \mathcal{L}_{\text{MaskedL1}} = \frac{\sum_{u,v} \mathcal{M}(u,v) \cdot |\hat{D}(u,v) - D(u,v)|}{\sum_{u,v} \mathcal{M}(u,v)}
    \end{equation}
    
    \item \textbf{Hybrid RMSE Loss:} Weighted combination of masked and global RMSE:
    \begin{equation}
        \mathcal{L}_{\text{HybridRMSE}} = \alpha \cdot \mathcal{L}_{\text{MaskedRMSE}} + (1-\alpha) \cdot \mathcal{L}_{\text{RMSE}}
    \end{equation}
    with $\alpha$ as a tunable parameter to emphasize object regions while maintaining weak global regularization.
    
    \item \textbf{Hybrid L1 Loss:} Weighted combination of masked and global L1:
    \begin{equation}
        \mathcal{L}_{\text{HybridL1}} = \alpha \cdot \mathcal{L}_{\text{MaskedL1}} + (1-\alpha) \cdot \mathcal{L}_{\text{L1}}
    \end{equation}
\end{enumerate}

The masked variants are designed to focus training explicitly on object geometry by excluding background pixels from loss computation, while hybrid losses balance object-focused learning with global regularization to prevent pathological solutions such as extreme scale drift.

\begin{figure}[ht]
    \centering
    \includegraphics[width=0.48\linewidth]{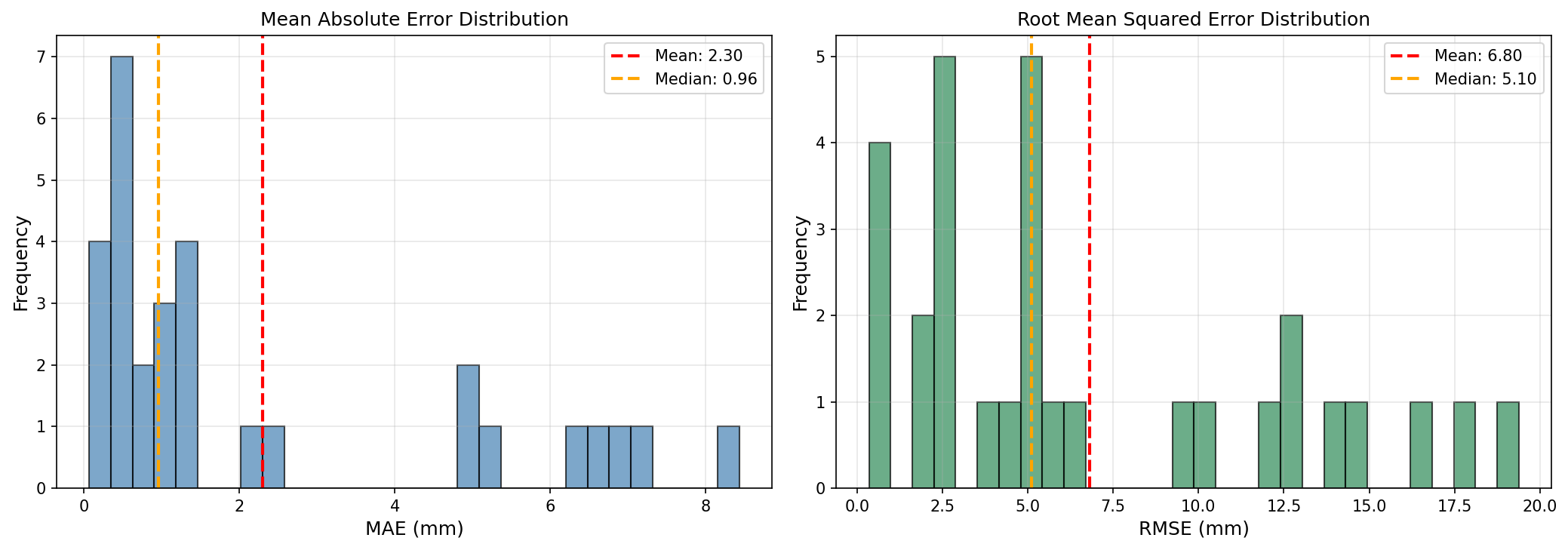}
    \includegraphics[width=0.48\linewidth]{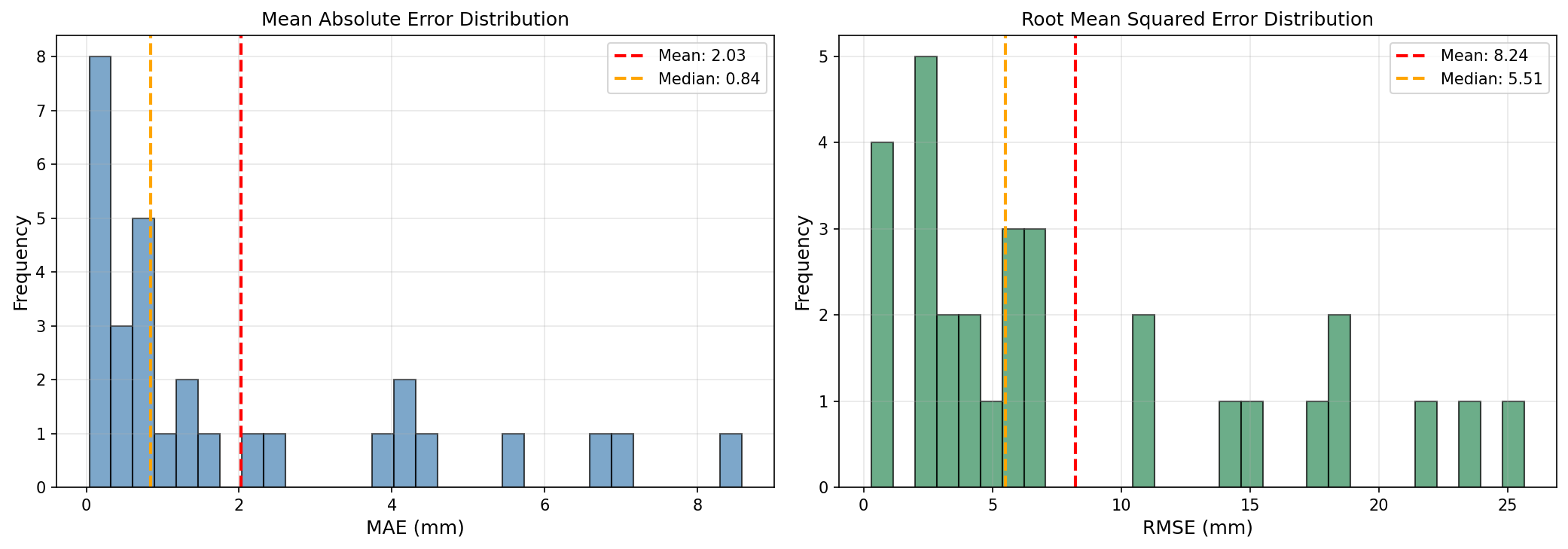}\\
    \vspace{0.2cm}
    \includegraphics[width=0.48\linewidth]{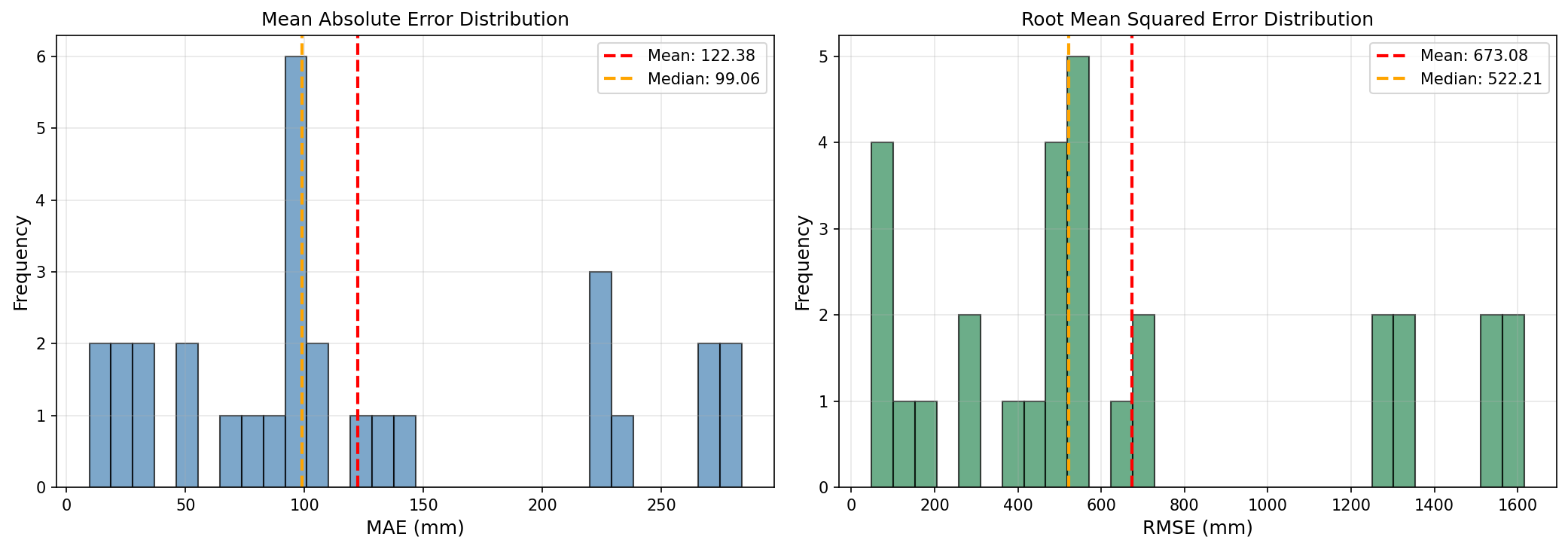}
    \includegraphics[width=0.48\linewidth]{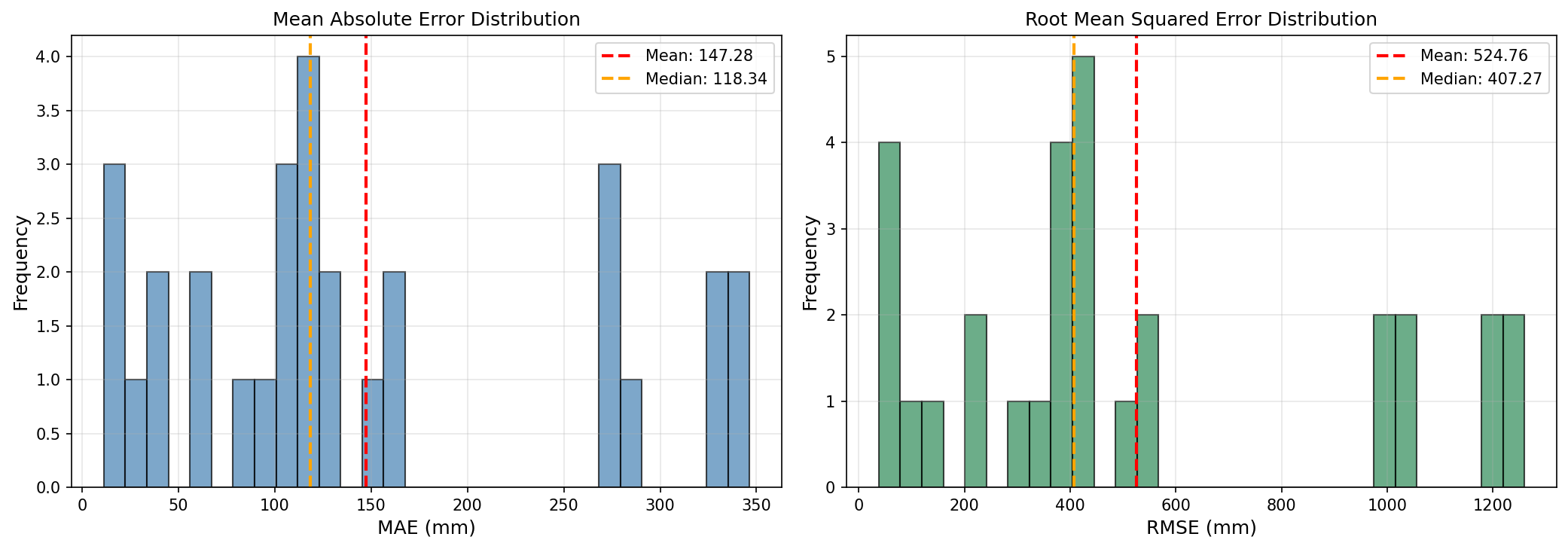}\\
    \vspace{0.2cm}
    \includegraphics[width=0.48\linewidth]{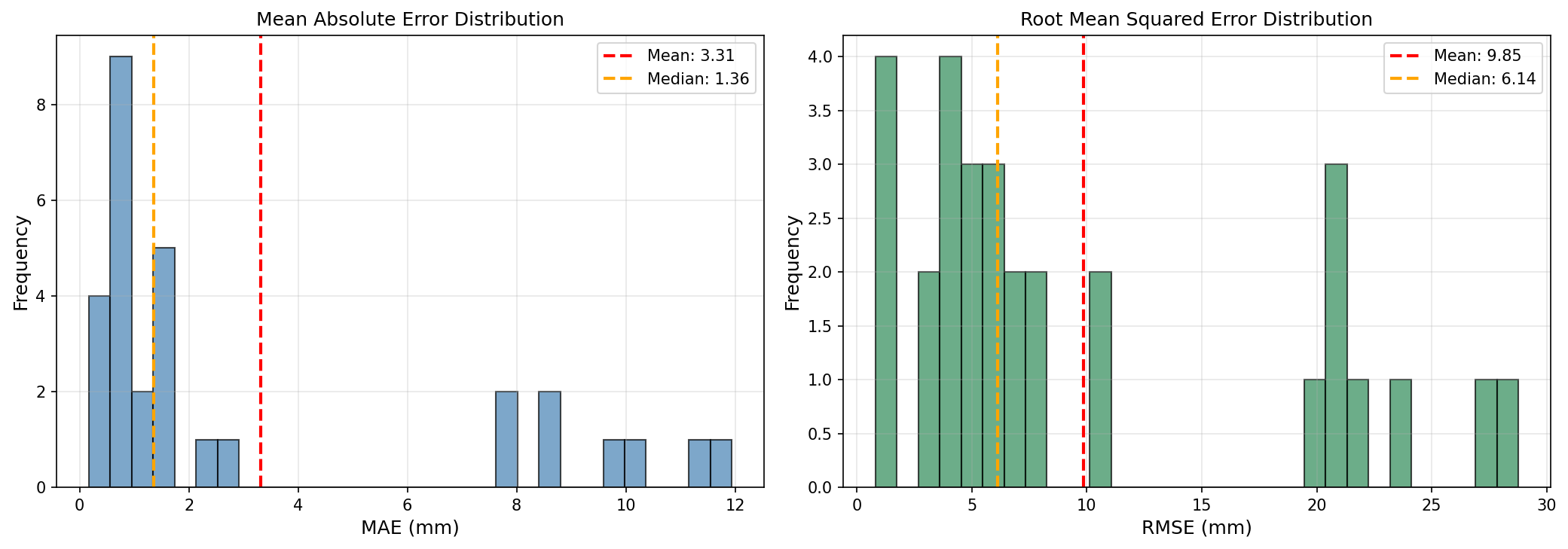}
    \includegraphics[width=0.48\linewidth]{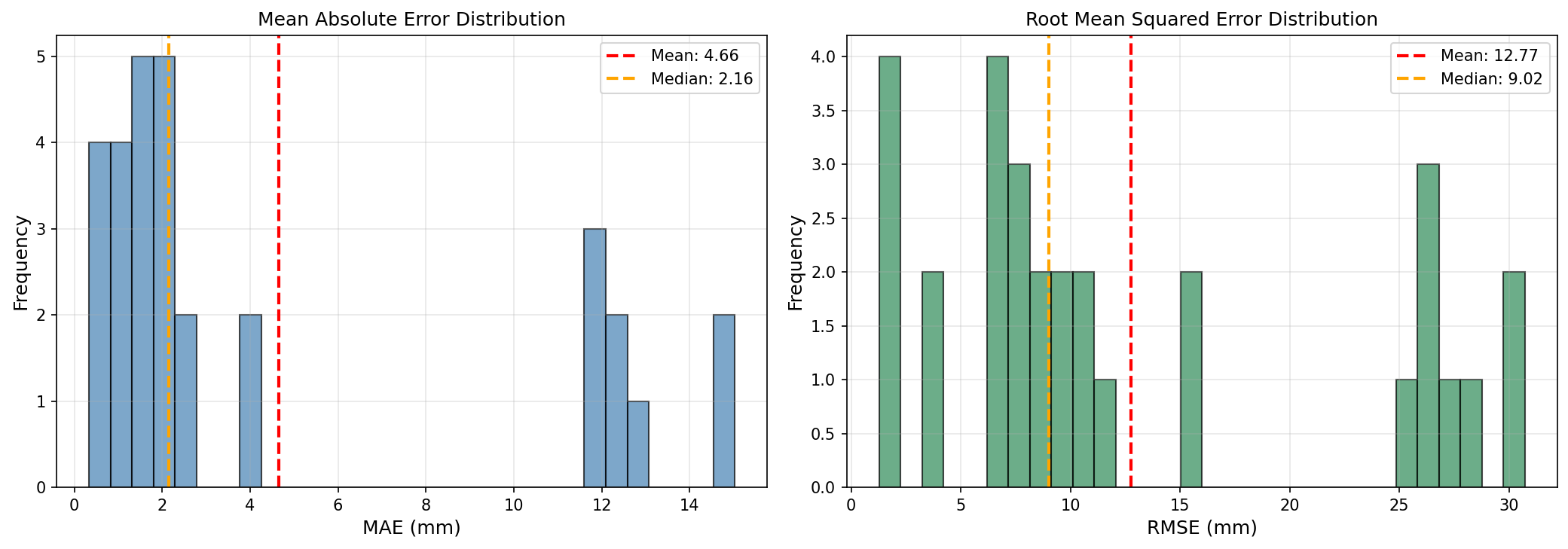}
    \caption{Overall error distributions for six loss functions on individual normalized depth. Top row: RMSE (2.30~mm MAE) and L1 (2.03~mm MAE) baselines. Middle row: Masked RMSE (122.38~mm MAE) and Masked L1 (147.28~mm MAE) exhibit catastrophic scale drift. Bottom row: Hybrid L1 with $\alpha=0.7$ (3.31~mm MAE, optimal) and $\alpha=0.9$ (4.66~mm MAE) balance object focus with scale stability.}
    \label{fig:loss_results}
\end{figure}

\begin{table}[ht]
\caption{Loss function comparison on individual normalized depth with full fringe inputs (30 test samples). All errors in millimeters.}
\label{tab:loss_comparison}
\begin{center}
\begin{tabular}{|l|c|c|c|c|c|c|}
\hline
\rule[-1ex]{0pt}{3.5ex} \textbf{Loss Function} & \textbf{Overall} & \textbf{Overall} & \textbf{Object} & \textbf{Object} & \textbf{Background} & \textbf{Background} \\
\rule[-1ex]{0pt}{3.5ex} & \textbf{MAE} & \textbf{RMSE} & \textbf{MAE} & \textbf{RMSE} & \textbf{MAE} & \textbf{RMSE} \\
\hline
\rule[-1ex]{0pt}{3.5ex} RMSE (Baseline) & 2.30 & 6.80 & 16.20 & 21.19 & \textbf{0.92} & \textbf{3.00} \\
\hline
\rule[-1ex]{0pt}{3.5ex} L1 & 2.03 & 8.24 & 19.34 & 25.93 & 0.13 & 2.61 \\
\hline
\rule[-1ex]{0pt}{3.5ex} Masked RMSE & 122.38 & 673.08 & 18.83 & 23.02 & 135.67 & 714.82 \\
\hline
\rule[-1ex]{0pt}{3.5ex} Masked L1 & 147.28 & 524.76 & 22.65 & 27.97 & 163.53 & 557.26 \\
\hline
\rule[-1ex]{0pt}{3.5ex} Hybrid RMSE ($\alpha$=0.5) & 2.91 & 7.36 & 15.29 & 19.40 & 1.64 & 4.80 \\
\hline
\rule[-1ex]{0pt}{3.5ex} Hybrid RMSE ($\alpha$=0.7) & 3.94 & 8.39 & 14.55 & 18.69 & 2.86 & 6.49 \\
\hline
\rule[-1ex]{0pt}{3.5ex} Hybrid RMSE ($\alpha$=0.9) & 6.00 & 10.92 & 15.05 & 19.19 & 5.24 & 9.83 \\
\hline
\rule[-1ex]{0pt}{3.5ex} Hybrid L1 ($\alpha$=0.5) & 2.90 & 8.99 & 15.41 & 19.37 & 1.53 & 7.09 \\
\hline
\rule[-1ex]{0pt}{3.5ex} Hybrid L1 ($\alpha$=0.7) & \textbf{3.31} & \textbf{9.85} & \textbf{14.54} & \textbf{17.88} & 2.01 & 8.44 \\
\hline
\rule[-1ex]{0pt}{3.5ex} Hybrid L1 ($\alpha$=0.9) & 4.66 & 12.77 & 14.73 & 18.85 & 3.63 & 11.93 \\
\hline
\end{tabular}
\end{center}
\end{table}

\begin{figure}[ht]
    \centering
    \includegraphics[width=0.95\linewidth]{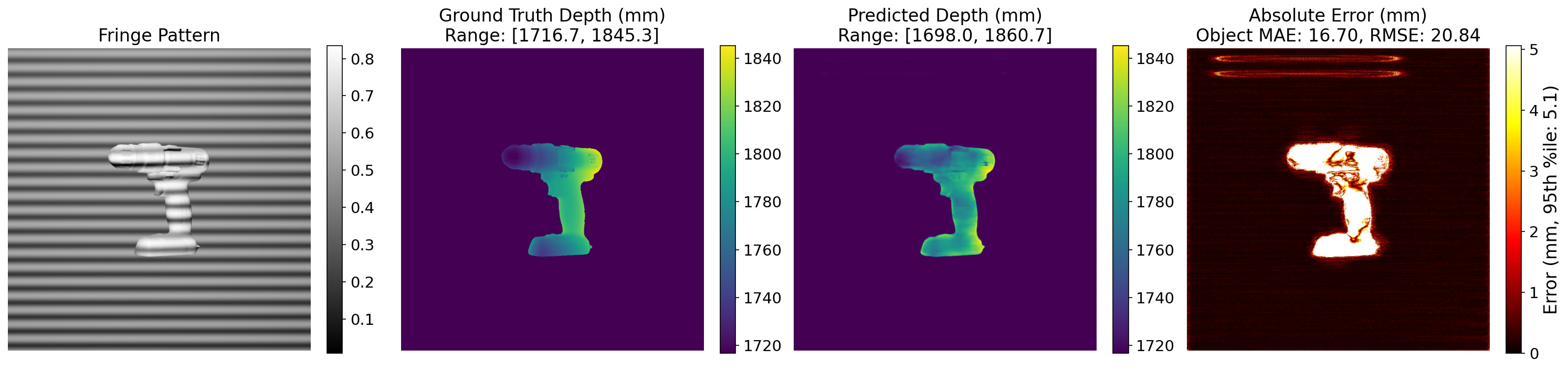}\\
    \vspace{0.2cm}
    \includegraphics[width=0.95\linewidth]{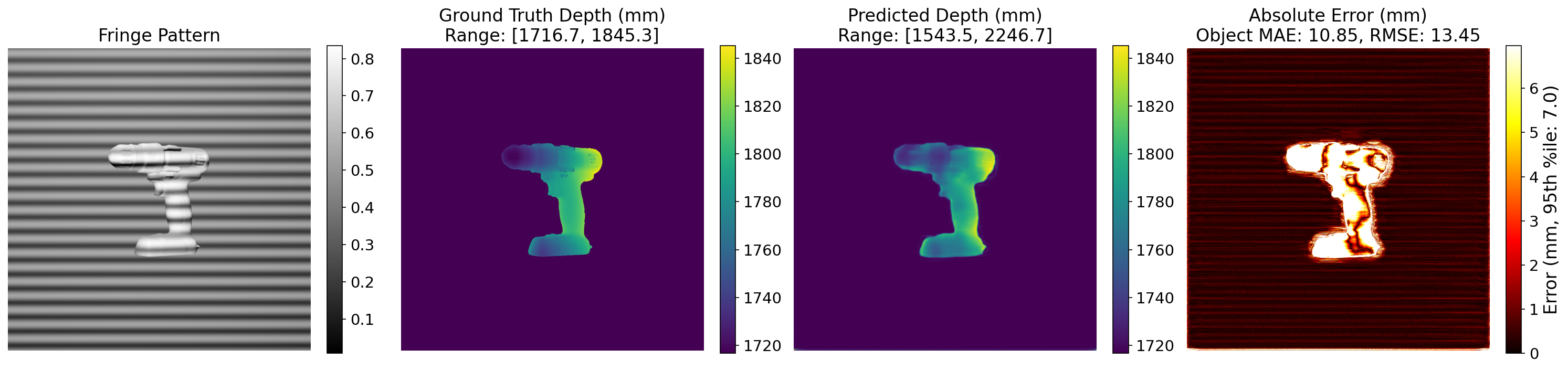}\\
    \vspace{0.2cm}
    \includegraphics[width=0.95\linewidth]{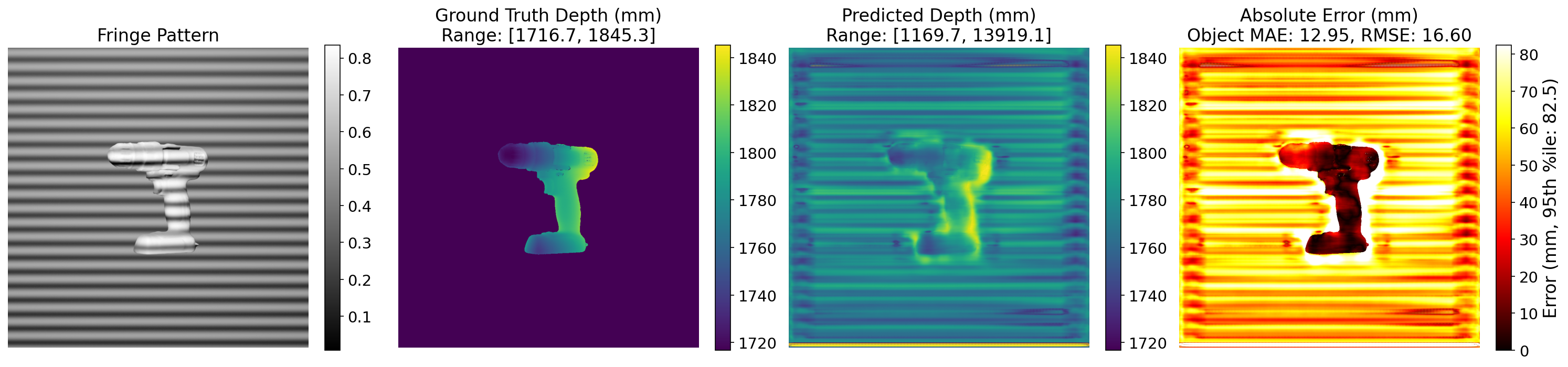}
    \caption{Single-shot depth reconstruction for power drill object under three representative loss functions. Top: RMSE baseline (16.70~mm object MAE). Middle: Hybrid L1 with $\alpha=0.7$ (12.21~mm object MAE, 27\% improvement, best overall). Bottom: Masked RMSE (23.93~mm object MAE) shows scale drift artifacts in background despite lower nominal object error.}
    \label{fig:loss_predictions}
\end{figure}

\textbf{Results:} Table~\ref{tab:loss_comparison} and Figure~\ref{fig:loss_results} summarize the results of six loss function families across varying weighting parameters. The results show three distinct behaviors based on how losses balance object and background pixels.

\textbf{Standard losses (RMSE, L1):} The baseline RMSE achieves excellent overall error (2.30~mm MAE, 6.80~mm RMSE) with superior background suppression (0.92~mm MAE) but moderate object error (16.20~mm MAE). L1 loss improves background suppression further (0.13~mm MAE) but increases object error to 19.34~mm (19\% worse), demonstrating the robustness-accuracy tradeoff inherent to L1 versus L2 objectives.

\textbf{Masked losses (Masked RMSE, Masked L1):} These losses catastrophically fail despite theoretically focusing on object geometry. Masked RMSE produces 122.38~mm overall MAE with extreme background errors (135.67~mm MAE, 714.82~mm RMSE), indicating severe scale drift. Figure~\ref{fig:loss_predictions} (bottom) shows horizontal banding artifacts characteristic of unconstrained scale. This failure occurs because masked losses provide no constraint on background predictions, allowing the network to converge to arbitrary offsets that minimize masked error at the expense of global consistency. The individual normalization's per-sample scale compounds this issue.

This failure mode directly parallels the background fringe ablation study (Section~\ref{sec:background_ablation}), where physically removing background fringes degraded performance 2.8-7.3×. Both failures stem from the same mechanism: \emph{removing spatial context}. Background fringe removal creates artificial black-to-object transitions and eliminates the continuous phase reference. Masked losses achieve a similar effect through gradient masking - by providing zero gradients on background pixels, the network learns to ignore background entirely, allowing predictions to drift to arbitrary scales that minimize object-only error. Both experiments demonstrate that background information, whether through direct fringe patterns or regularization constraints, is essential for stable depth prediction in single-shot FPP.

\textbf{Hybrid losses ($\alpha$ ablation):} To balance object-focused learning with scale stability, we evaluated hybrid losses across three weighting values: $\alpha=0.5$ (equal weighting), $\alpha=0.7$ (moderate object emphasis), and $\alpha=0.9$ (strong object emphasis). Hybrid L1 with $\alpha=0.7$ achieves the best object-only performance (14.54~mm MAE, 17.88~mm RMSE), representing a 10\% improvement over baseline RMSE (16.20~mm MAE) and 25\% improvement over L1 (19.34~mm MAE). Figure~\ref{fig:loss_predictions} (middle) shows clean predictions without the scale drift artifacts of masked losses. Increasing $\alpha$ to 0.9 slightly degrades performance (14.73~mm object MAE), while reducing to 0.5 approaches baseline performance (15.41~mm MAE), demonstrating an inverted-U relationship where moderate object emphasis is optimal. The weak global regularization term $(1-\alpha) \cdot \mathcal{L}_{\text{global}}$ prevents pathological solutions while the dominant masked term $\alpha \cdot \mathcal{L}_{\text{masked}}$ focuses training on object geometry.

The modest improvements from hybrid losses (10-16\% object MAE reduction) suggest that the baseline RMSE is already near-optimal for the information available in single fringe images. The periodic nature of fringe patterns and absence of temporal unwrapping fundamentally limit achievable accuracy, and loss function engineering provides only marginal gains. These results establish Hybrid L1 with $\alpha$=0.7 as the best-performing configuration for subsequent architecture comparison.

\subsection{Architecture Comparison}\label{sec:models}

Having identified individual normalization and Hybrid L1 loss ($\alpha$=0.7) as the optimal configuration, we now benchmark this setup across four representative architectures to evaluate whether model design can overcome the fundamental information limitations of single-shot reconstruction. The optimizer and other hyperparameters were kept the same as mentioned in Section~\ref{sec:normalization_comparison}.

\textbf{UNet}~\cite{ronneberger2015}: Encoder-decoder with skip connections (described in Section~\ref{sec:normalization_comparison}).

\textbf{Hformer}~\cite{zhu2022hformer}: Hybrid CNN-transformer with HRNet-W18 backbone for multi-scale features [18,36,72,144], transformer encoder-decoder with window-based attention (size 8), patch expansion upsampling. 

\textbf{ResUNet}~\cite{ikeda2025deep}: UNet with residual blocks replacing convolutional blocks, four levels ($960\times960$ to $120\times120$), identity skip connections for improved gradient flow. 

\textbf{Pix2Pix}~\cite{isola2017}: Conditional GAN with U-Net generator and PatchGAN discriminator, adapted from NVIDIA Pix2Pix-HD~\cite{wang2018}.

\begin{figure}[ht]
    \centering
    \includegraphics[width=0.48\linewidth]{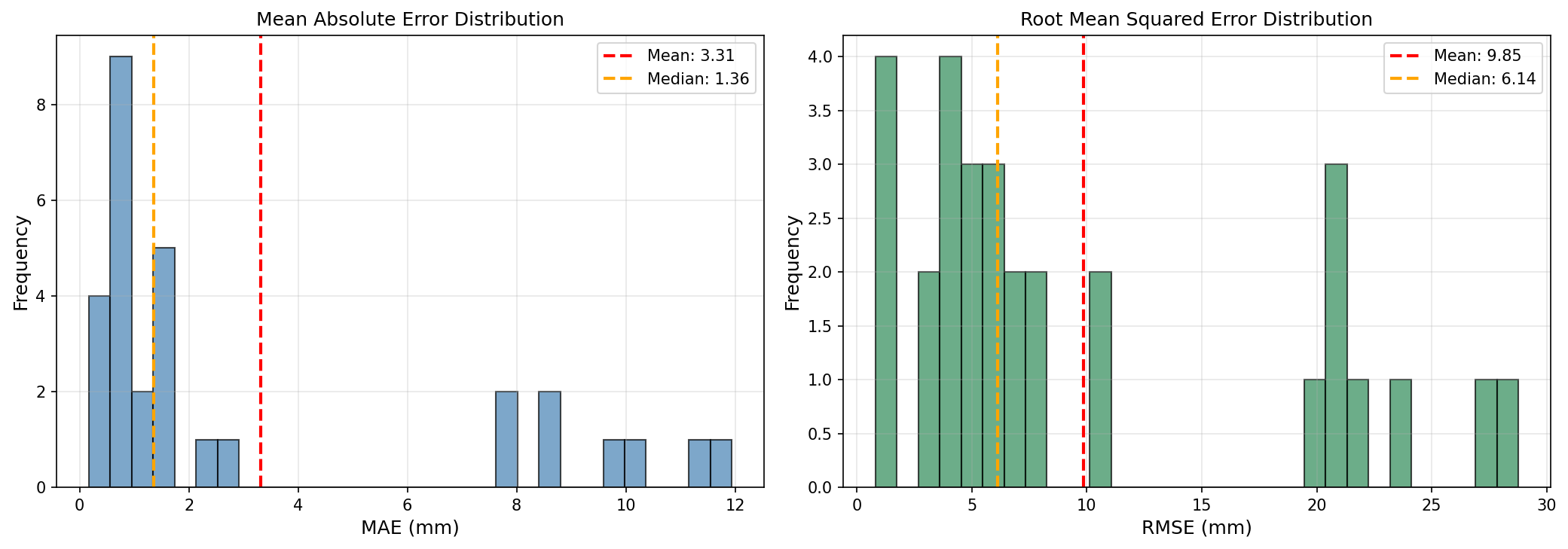}
    \includegraphics[width=0.48\linewidth]{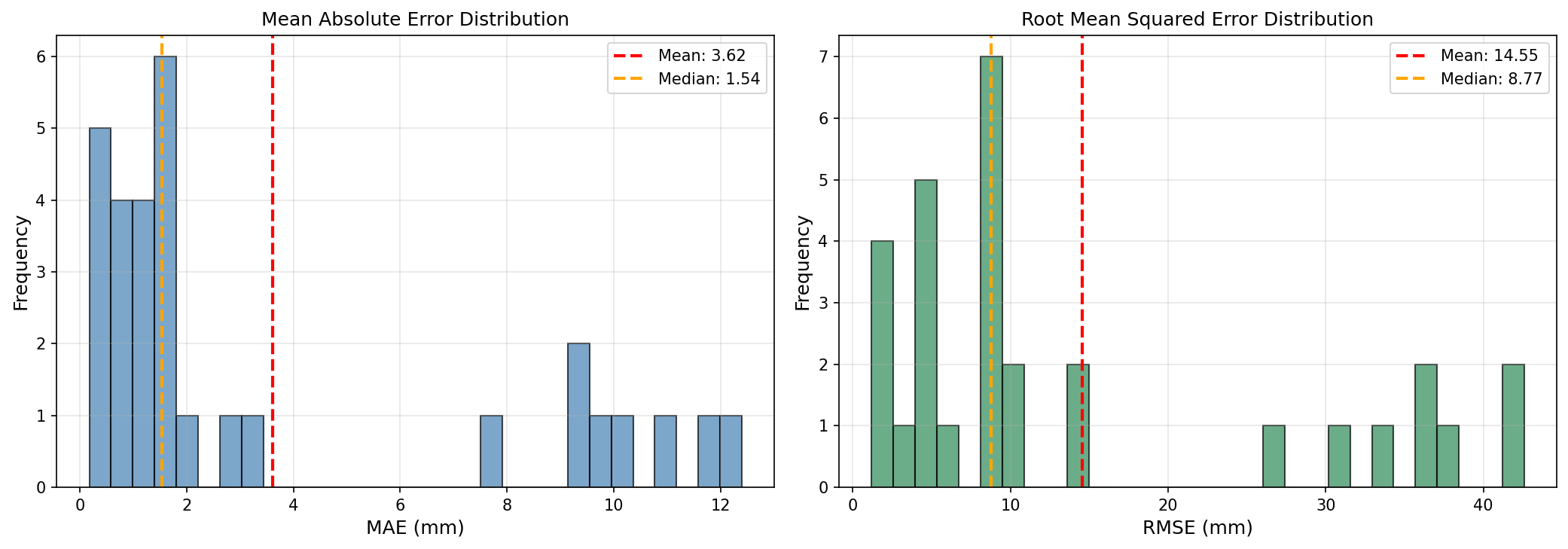}\\
    \vspace{0.2cm}
    \includegraphics[width=0.48\linewidth]{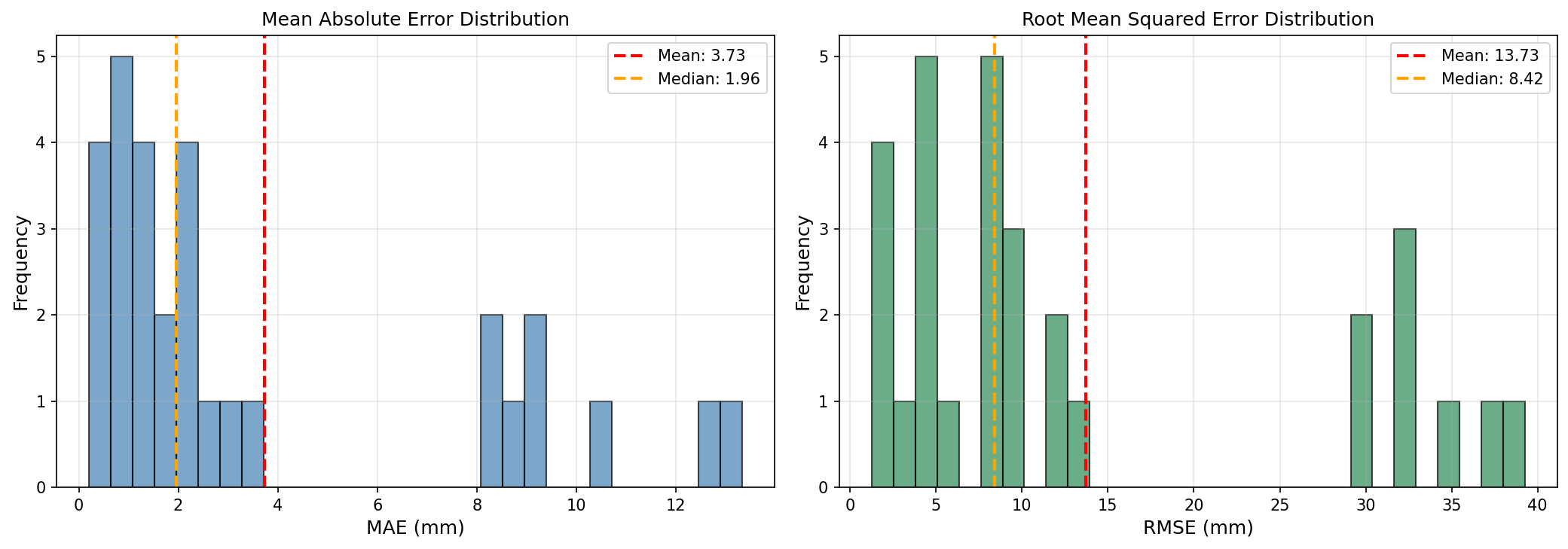}
    \includegraphics[width=0.48\linewidth]{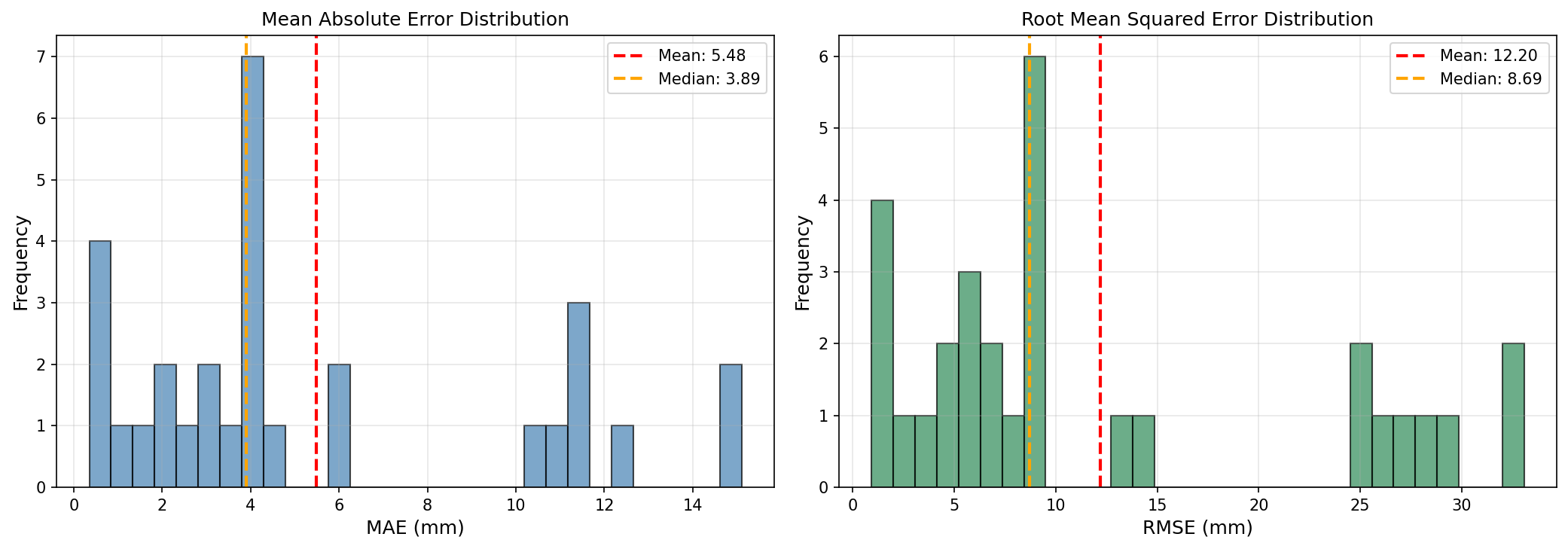}
    \caption{Overall error distributions for four architectures trained with optimal configuration (individual normalization + Hybrid L1 $\alpha$=0.7). Top row: UNet (3.31~mm overall MAE, best) and Hformer (3.62~mm overall MAE). Bottom row: ResUNet (3.73~mm overall MAE) and Pix2Pix (5.48~mm overall MAE, worst). UNet shows tightest distribution with lowest variance.}
    \label{fig:architecture_results}
\end{figure}

\begin{table}[ht]
\caption{Architecture comparison on individual normalized depth with Hybrid L1 loss ($\alpha$=0.7) across 30 test samples. All errors in millimeters. UNet achieves best performance, but all models show substantial object-only errors.}
\label{tab:architecture_comparison}
\begin{center}
\begin{tabular}{|l|c|c|c|c|c|c|}
\hline
\rule[-1ex]{0pt}{3.5ex} \textbf{Architecture} & \textbf{Overall} & \textbf{Overall} & \textbf{Object} & \textbf{Object} & \textbf{Background} & \textbf{Background} \\
\rule[-1ex]{0pt}{3.5ex} & \textbf{MAE} & \textbf{RMSE} & \textbf{MAE} & \textbf{RMSE} & \textbf{MAE} & \textbf{RMSE} \\
\hline
\rule[-1ex]{0pt}{3.5ex} UNet & \textbf{3.31} & \textbf{9.85} & \textbf{14.54} & \textbf{17.88} & \textbf{2.01} & \textbf{8.44} \\
\hline
\rule[-1ex]{0pt}{3.5ex} Hformer & 3.62 & 14.55 & 22.17 & 26.78 & 1.72 & 12.87 \\
\hline
\rule[-1ex]{0pt}{3.5ex} ResUNet & 3.73 & 13.73 & 23.32 & 28.38 & 1.80 & 11.54 \\
\hline
\rule[-1ex]{0pt}{3.5ex} Pix2Pix & 5.48 & 12.20 & 27.73 & 38.22 & 3.16 & 5.15 \\
\hline
\end{tabular}
\end{center}
\end{table}

\begin{figure}[ht]
    \centering
    \includegraphics[width=0.95\linewidth]{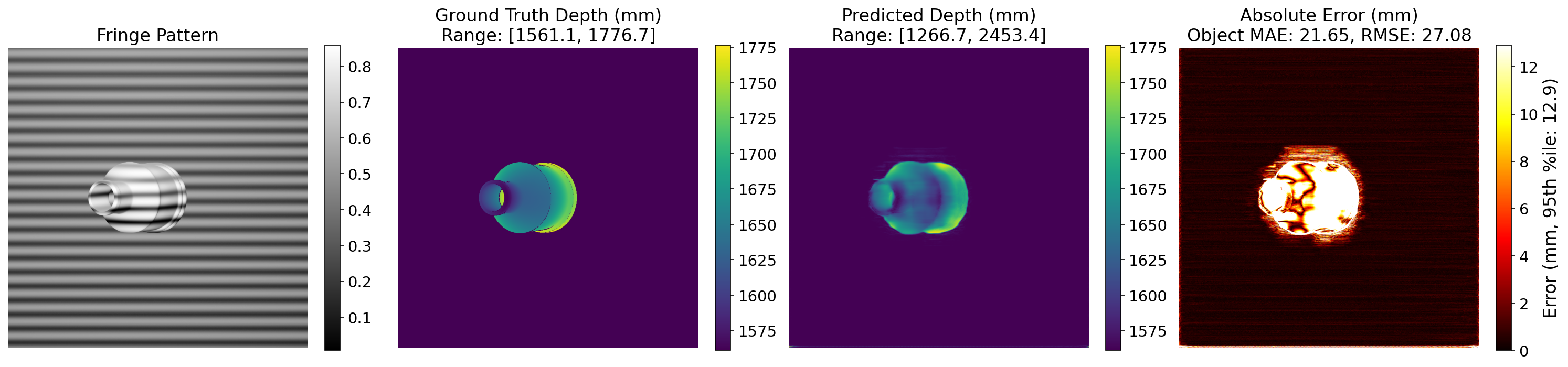}\\
    \vspace{0.2cm}
    \includegraphics[width=0.95\linewidth]{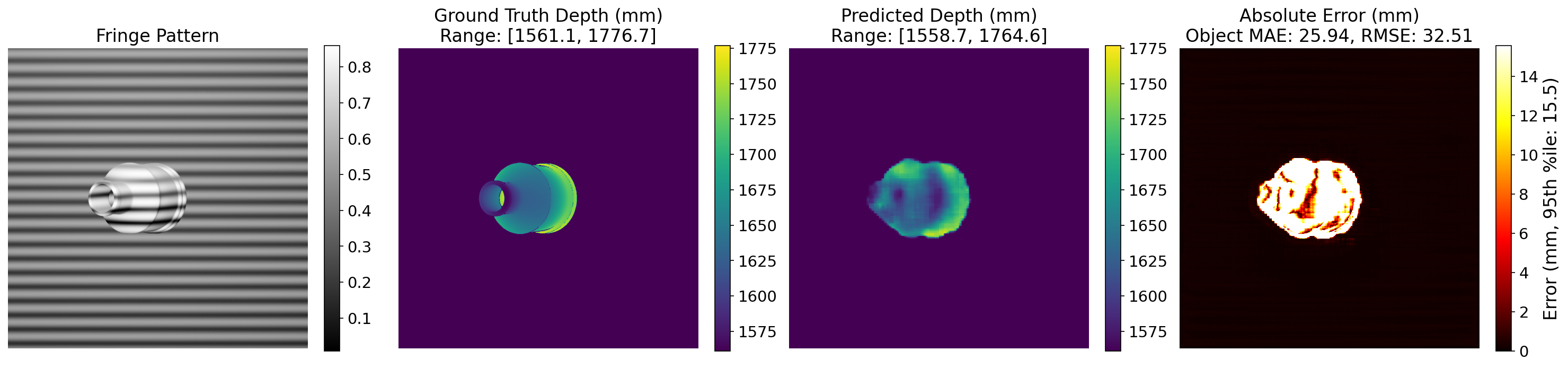}\\
    \vspace{0.2cm}
    \includegraphics[width=0.95\linewidth]{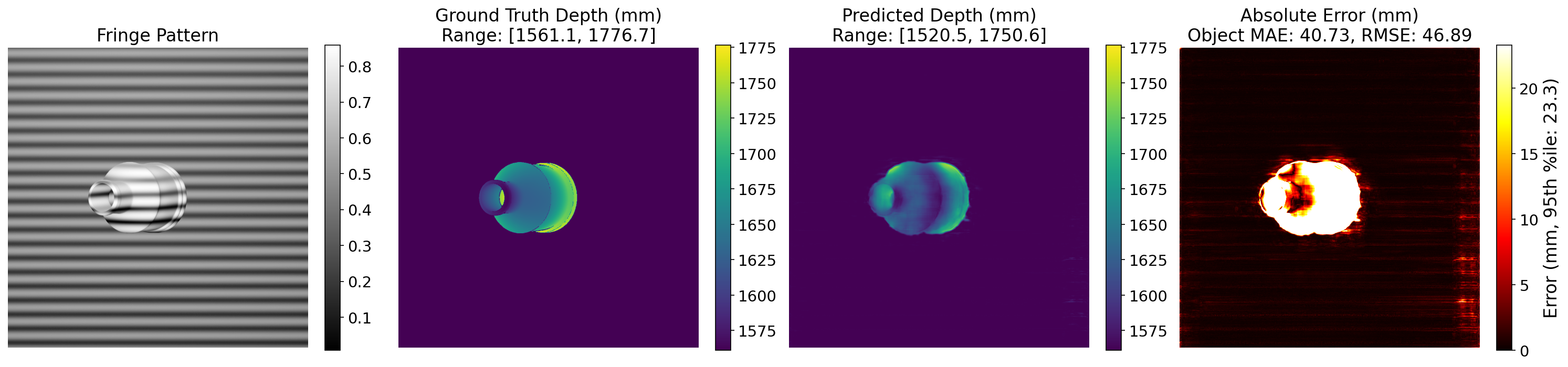}\\
    \vspace{0.2cm}
    \includegraphics[width=0.95\linewidth]{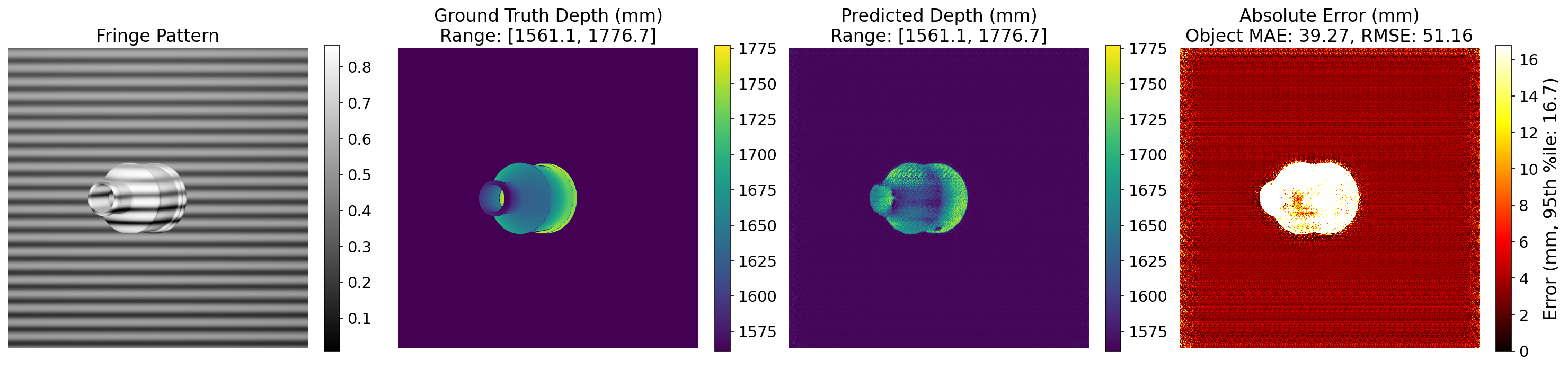}
    \caption{Single-shot depth reconstruction for container bottle object across four architectures. From top to bottom: UNet (21.65~mm object MAE, 27.08~mm object RMSE), Hformer (25.94~mm object MAE, 32.51~mm object RMSE), ResUNet (40.73~mm object MAE, 46.89~mm object RMSE), Pix2Pix (39.27~mm object MAE, 51.16~mm object RMSE, worst performance). All models capture coarse geometry but fail on fine-scale accuracy. Note severe background artifacts in Pix2Pix prediction with uniformly high error across the entire background region, characteristic of adversarial training's focus on perceptual quality over metric accuracy.}
    \label{fig:architecture_predictions}
\end{figure}

\textbf{Results:} Table~\ref{tab:architecture_comparison} and Figure~\ref{fig:architecture_results} summarize performance across architectures. UNet achieves the best object-only error (14.54~mm MAE, 17.88~mm RMSE), outperforming Hformer by 52\%, ResUNet by 60\%, and Pix2Pix by 91\%. Figure~\ref{fig:architecture_predictions} shows representative predictions for a container bottle object.

UNet's superior performance under the optimized configuration suggests that the baseline encoder-decoder architecture with skip connections is well-suited for this task when combined with proper normalization and loss function design. The consistent [0,1] normalized space and hybrid loss allow UNet's simple architecture to focus learning on shape reconstruction without the complexities of handling scale variation or architectural sophistication.

Hformer and ResUNet achieve similar mid-range performance (22-23~mm object MAE), despite their architectural differences. Hformer's transformer-based attention mechanisms and ResUNet's residual connections provide comparable benefits, but neither architecture overcomes the fundamental information deficit of single-shot reconstruction. The 52-60\% degradation compared to UNet suggests that architectural complexity may actually hinder performance on this task, potentially due to overfitting on the limited 240-sample training set or difficulty optimizing more complex models in the constrained [0,1] normalized space.

Pix2Pix performs worst (27.73~mm object MAE, 38.22~mm object RMSE), revealing a fundamental misalignment between adversarial training and metrological objectives. Figure~\ref{fig:architecture_predictions} shows a counterintuitive result: Pix2Pix produces visually compelling predictions with smooth surfaces and statistically correct depth distributions---the container bottle prediction exhibits a depth range [1561.1, 1776.7]~mm that exactly matches ground truth---yet achieves nearly 2$\times$ worse error than UNet (14.54~mm MAE). This occurs because the adversarial discriminator optimizes for perceptual plausibility (``does this \emph{look} like a depth map?'') rather than point-wise accuracy. The network learns to match statistical properties (min, max, mean depth) and produce realistic surface geometries, but introduces smooth systematic offsets of 20-40~mm that are visually imperceptible yet catastrophic for precision measurement. These offsets appear as uniform color shifts in visualizations rather than high-frequency noise, making predictions appear deceptively accurate to human observers while failing metrological evaluation. The severe background artifacts further confirm this: the network learned that ``depth maps should have varying depth values'' without the critical constraint that background regions must predict exactly zero depth.

Critically, even the best architecture (UNet at 14.54~mm MAE) remains far from the sub-millimeter accuracy of traditional FPP. For objects with an 80~mm depth range in our dataset, this represents 18\% relative error. The modest 1.9$\times$ performance gap between best (UNet: 14.54~mm) and worst (Pix2Pix: 27.73~mm) architectures, combined with all models' failure to achieve precision depth reconstruction, confirms our hypothesis: the fundamental limitation is information deficit, not model design. Single fringe images lack sufficient information for accurate depth recovery, and no amount of architectural engineering can overcome this constraint.

This finding parallels the loss function results (Section~\ref{sec:loss_comparison}), where even optimal loss design provided only 10\% improvement over baseline. Together, these results demonstrate that single-shot fringe-to-depth mapping without explicit phase information is fundamentally limited. Networks learn coarse shape priors and statistical regularities rather than accurate geometry, regardless of architectural sophistication.

\section{Conclusion and Future Work}
\label{sec:conclusion}

This paper presents the first comprehensive open-source machine learning benchmarking framework for fringe projection profilometry, establishing both optimal learning configurations and fundamental limitations for single-shot 3D reconstruction. Using a large-scale synthetic dataset (15,600 fringe images, 300 depth reconstructions, 50 objects) generated with VIRTUS-FPP's physics-based rendering. We conducted a systematic three-phase ablation study that reveals critical insights for learning-based FPP.

\textbf{Phase 1 - Data representation}: Individual normalization, which decouples object shape learning from absolute scale, achieves 9.1$\times$ better reconstruction (16.20~mm vs 148.07~mm object MAE) than raw depth and 5.1$\times$ better than global normalization. Unexpectedly, removing background fringe patterns degrades performance 2.8-7.3$\times$ across all normalizations, demonstrating that background fringes provide essential spatial phase reference rather than acting as noise. This finding challenges conventional assumptions about input preprocessing for FPP learning.

\textbf{Phase 2 - Loss function optimization}: Among six L1/L2-based losses, Hybrid L1 with $\alpha$=0.7 achieves best performance (14.54~mm object MAE, 10\% improvement over baseline). Critically, masked losses that focus exclusively on object pixels catastrophically fail (122-147~mm overall MAE) due to scale drift, paralleling the background ablation finding: both removing background fringes physically and masking them computationally eliminate essential spatial constraints, implying background information is signal, not noise, for single-shot FPP.

\textbf{Phase 3 - Architecture comparison}: Using the obtained optimal configuration (individual normalization + Hybrid L1 $\alpha$=0.7), UNet achieves best performance (14.54~mm object MAE, 17.88~mm object RMSE), outperforming Hformer by 52\%, ResUNet by 60\%, and Pix2Pix by 91\%. The modest 1.9$\times$ gap between best (14.54~mm) and worst (27.73~mm) architectures, combined with all models' failure to achieve sub-millimeter accuracy, confirms that information deficit limits reconstruction quality. Pix2Pix's counterintuitive failure reveals that its predictions exhibit statistically correct depth distributions (e.g., exact range match [1561-1777~mm]) and smooth, realistic surfaces, yet achieve worst metric accuracy. This reveals fundamental misalignment between adversarial training's perceptual plausibility objective and metrology's point-wise accuracy requirement, explaining why modern depth estimation do not use GANs.

Even the best model's 14.54~mm error represents 18\% of objects in our dataset with 80~mm depth range, far from traditional FPP's sub-millimeter capabilities. This demonstrates that single fringe images fundamentally lack information for accurate depth recovery without explicit phase: the periodic nature of sinusoidal patterns creates inherent $2\pi$ ambiguity that cannot be resolved through learned shape priors alone. Networks achieve coarse semantic reconstruction but fail on precise geometry.

These findings strongly motivate hybrid approaches combining traditional phase-based FPP with learned refinement. Future directions include: (1) \textbf{Phase-guided learning} using wrapped/unwrapped phase maps as network input or intermediate representations, (2) \textbf{Multi-view fusion} leveraging 6 viewpoints per object in our dataset, (3) \textbf{Sim-to-real transfer} via domain adaptation using VIRTUS-FPP's digital twin capability, (4) \textbf{Task reformulation} for post-processing traditional reconstructions (denoising, hole-filling, outlier removal) rather than end-to-end depth prediction, (5) \textbf{Dataset expansion} to challenging materials (specular, translucent) and lighting conditions with domain randomization~\cite{tobin2017domain}, and (6) \textbf{Uncertainty quantification} through probabilistic deep learning to identify when predictions are unreliable.

By providing a comprehensive open-source synthetic data, standardized evaluation protocols separating object from background errors, and systematic evidence of fundamental limitations, this work establishes a foundation for data-driven FPP research. Our findings suggest that rather than pursuing increasingly complex architectures for direct fringe-to-depth mapping, the field should focus on integrating learned components with physics-based reconstruction pipelines to leverage both traditional FPP's geometric precision and neural networks' robustness to noise and ambiguity.





\acknowledgments

We thank Iowa State University for access to computational resources.

\bibliography{report}
\bibliographystyle{spiebib}

\end{document}